\documentclass[prd,preprint,superscriptaddress,amsmath,amssymb,nofootinbib]{revtex4}
\usepackage{graphicx}
\usepackage{dcolumn}
\usepackage{bm}
\usepackage{amssymb}
\usepackage{amsmath}
\usepackage{epsfig}    
\usepackage{color}
\usepackage{slashed}
\usepackage{hhline}

\def\be{\begin{equation}}
\def\ee{\end{equation}}
\newcommand{\bea}{\begin{eqnarray}}
\newcommand{\eea}{\end{eqnarray}}
\newcommand{\nn}{\nonumber}

\begin{document}
\title{A radiatively induced inverse seesaw model \\
 with hidden $U(1)$ gauge symmetry}
\preprint{CTP-SCU/2021009, APCTP Pre2021-005}


\author{Takaaki Nomura}
\email{nomura@scu.edu.cn}
\affiliation{College of Physics, Sichuan University, Chengdu 610065, China}

\author{Hiroshi Okada}
\email{hiroshi.okada@apctp.org}
\affiliation{Asia Pacific Center for Theoretical Physics (APCTP) - Headquarters San 31, Hyoja-dong,
Nam-gu, Pohang 790-784, Korea}

\author{Prasenjit Sanyal}
\email{ prasenjit.sanyal@apctp.org}
\affiliation{Asia Pacific Center for Theoretical Physics (APCTP) - Headquarters San 31, Hyoja-dong,
Nam-gu, Pohang 790-784, Korea}

\date{\today}

\begin{abstract}
We propose an inverse seesaw scenario under hidden $U(1)$ gauge symmetry, having rather natural hierarchies among neutral fermion mass scales.
The hierarchies are derived from theory and experimental constraints. The theory requires Majorana exotic masses have to be induced at one-loop level. 
The experimental side requests that the Dirac mass terms has to be highly suppressed to satisfy the constraints from the lepton flavor violations(LFVs) such as $\mu\to e\gamma$. In order to induce such a small Majorana exotic masses we introduce exotic fermions and bosons that also provide us additional intriguing phenomenologies such as muon anomalous magnetic moment, dark matter candidate as well as LFVs and deviations in leptonic $Z$ decays. We analyze these phenomenologies including neutrino oscillation data numerically, and show allowed region. 
Finally, we discuss the DM candidate in both the cases of fermion and scalar boson,
where we focus on rather lighter range that is equal or less than 10 GeV. The dominant contributions originate from interactions of hidden gauge sector, and we show allowed ranges for both cases.
\end{abstract}
\maketitle

\section{Introduction}
Understanding electrically neutral fields is our common issues to be resolved in both theoretical and experimental view points. 
Typically, it is considered in the framework of physics beyond the standard model (SM) since the active neutrino masses as well as dark matter(DM) candidate cannot be explained in the SM unless new scale or fields are supplied. 
Even though several experiments have partially discovered nature of the active neutrinos measuring three mixing angles and two squared mass differences, we only have few ideas about CP phases etc.; Dirac CP phase, Majorana phases, neutrino mass eigenvalues,  Majorana or Dirac mass type, and so on.
This is because the mechanism to generate neutrino masses is not uniquely determined yet. 
On the other hand there are almost null direct information on nature of DM in spite of huge efforts by experimentalists on direct searches such as XENON1T~\cite{Aprile:2018dbl} and LUX~\cite{Akerib:2016vxi}, indirect searches such as Fermi-LAT~\cite{FermiLAT:2011ab}, AMS-02~\cite{Aguilar:2013qda}, and CALET~\cite{Adriani:2017efm}, and collider searches at LHC~\cite{Khachatryan:2014rra}. 
These experiments especially focus on weakly interacting massive particles (WIMP) whose typical mass would be the order 100 GeV to 1000 GeV. 
Although it does not suggest WIMP has totally been excluded, it would be worthwhile to consider a lighter DM candidate below e.g., 10 GeV.  It would be more natural in view of relaxed hierarchies among neutral fields, since the active neutrino masses are minuscule  that is the order $10^{-9}$ GeV.
In order to realize such a tiny mass of neutrinos, we propose an improved inverse seesaw model with a hidden gauged $U(1)$ symmetry that plays crucial roles in carrying out our neutrino model and assuring the stability of our DM due to remnant $Z_2$ symmetry after spontaneous symmetry breaking.
Hidden $U(1)$ symmetry has an advantage that the mass of hidden gauge boson $(m_{Z'})$ and its gauge coupling $(g'')$ are almost taken to be free, since the $Z'$ boson does not directly couple to the SM fields and there are almost no bounds by LEP~\cite{Barate:2003sz} and LHC~\cite{Tanabashi:2018oca}.~\footnote{Even though a bound arises from kinetic mixing, one can arbitrary neglect this effect.}
It is unlikely to flavorful gauge symmetries such as $U(1)_{B-L}$. 
This is why this kind of symmetry is frequently applied to lighter DM scenarios~\cite{Zhang:2009dd, Chiang:2013kqa, Chen:2015nea, Chen:2015dea, Gross:2015cwa, Hambye:2008bq, Boehm:2014bia, Baek:2013dwa, Khoze:2014woa, Daido:2019tbm, Karam:2015jta, Davoudiasl:2013jma, Ko:2020qlt, Nomura:2020zlm, Cai:2018upp, Nomura:2017wxf}.

 The original inverse seesaw mass matrix in basis of $[\nu_L,N_R^C,N_L]^T$~\cite{Mohapatra:1986bd, Wyler:1982dd} can be sketched as follows:
\begin{align}
\left( \begin{array}{ccc} 0 & M_D & 0 \\ M_D &0 & M_N \\ 0 & M_N & \mu \end{array} \right),
\end{align}
then the active neutrino mass matrix is given by $\mu\left(\frac{M_D}{M_N}\right)^2$ after the block-diagonalization.
Notice here that it requires mass hierarchies $\mu< M_D<M_N$ at least in order to generate the appropriate neutrino mass scale. Usually, these mass hierarchies are imposed by hand.

To achieve the mass hierarchies, one of the interesting scenario is generation of Majorana $\mu$ mass term at loop level; here we call it as radiative inverse seesaw.
In ref.~\cite{Ma:2009gu}, a radiative inverse seesaw model is proposed applying local $U(1)_\chi$ motivated by $SO(10)$ and adding some scalar fields; 
a model with local $U(1)_{B-L}$ is found in ref.~\cite{Mandal:2019oth}.
Moreover a model with lepton triplet~\cite{Law:2012mj} is considered where $\mu$ mass term at tree level is forbidden by the nature of gauge symmetry.
$Z_4$ symmetry can be also used to forbid tree level $\mu$ mass term~\cite{Ahriche:2016acx}.
A radiative inverse seesaw model with a new gauge symmetry is phenomenologically interesting since it provides new phenomena at experiments
~\footnote{Tree level inverse seesaw models with local/global $U(1)$ symmetries are also discussed by the current authors~\cite{Nomura:2018mwr,Dey:2019cts,Cai:2018upp}. Phenomenological constraints are also discussed in ref.~\cite{Das:2019pua} for general $U(1)$ gauge symmetry where the SM fermions are charged under it.}.
Here we are interested in a scenario with hidden $U(1)$ gauge symmetry where the SM fields are not charged under it.
A $Z'$ boson from hidden $U(1)$ would provide different phenomenology from non-hidden $U(1)$ symmetry such as $U(1)_{B-L}$.
Remarkably $Z'$ can have interactions with lepton flavor violation(LFV) when the SM lepton and exotic fermions with hidden $U(1)$ charge mixes.
It is thus interesting to consider a radiative inverse seesaw model with hidden $U(1)$ gauge symmetry and discuss various lepton flavor physics.

In this paper we propose a radiative inverse seesaw model with hidden $U(1)$ gauge symmetry.
In our model, we explain the hierarchies among $\{ \mu, M_D, M_N\}$ through theoretical and experimental manners;
$\mu$ is suppressed by generating it at one-loop level~\footnote{Similar models are found in e.g. refs.~\cite{Das:2017ski, Pilaftsis:1991ug, Dev:2012sg, Dev:2012bd}}, while $M_D/M_N$ has to be tiny enough to satisfy lepton flavor violations such as $\mu\to e\gamma$ process, as can be seen in our analysis.
Here $M_N$ is introduced as a mass of vector-like lepton doublet $L'$ and it is not suppressed.
Majorana mass $\mu$ is forbidden at tree level due to the nature of gauge symmetry. We realize it at one loop level where we introduce singlet fermion $N$,
 inert scalar doublet $\eta$ and singlet $\chi$ with nonzero hidden charges, in order to induce relevant loop diagrams. 
Here $\{N, \eta, \chi\}$ fields will have remnant $Z_2$ odd parity after spontaneous symmetry breaking of hidden $U(1)$ and the lightest neutral particle among them can be our DM candidate. 
 Thanks to these additional fields, we can also discuss flavor physics such as muon anomalous magnetic moment(muon $g-2$), LFVs and deviations in leptonic $Z$ boson decays
as well as DM phenomenologies. 

This paper is organized as follows.
In Sec.~II, we review our model and formulate the Higgs sector, and gauge sector.
In Sect.~III, we discuss the neutrino masses, LFVs, muon $g-2$, leptonic $Z$ boson decays, and show numerical analysis satisfying all the constraints except DM.
In Sect.~IV, we  study two types of our DM; bosonic and fermionic one, and demonstrate the allowed region to satisfy the relic density for each candidate. 
Notice here that the dominant processes are via new gauge interactions that would not be independent of neutrino and flavor physics.
In Sec.~V, we summarize and conclude.

\section{model}

\begin{table}[h!]
  \centering
  \begin{tabular}{|c|c|c|c|c|c|c|c||c|c|c|c|c|} \hline
   Fields  & ~$Q_L$~ & ~$U_R$~ & ~$d_R$~ & ~$L_L$~ & ~$e_R$~ & ~$L'_{L,R}$~  & ~$N_{L,R}$~ & ~$\Phi$~ & ~$\eta$~ & ~$\varphi$~ & ~$\chi$~ \\ 
    \hline
    $SU(3)_{C}$ & $\bm{3}$ & $\bm{3}$ & $\bm{3}$ & $\bm{1}$ & $\bm{1}$ & $\bm{1}$ & $\bm{1}$ & $\bm{1}$ & $\bm{1}$ & $\bm{1}$  & $\bm{1}$   \\
    \hline
    $SU(2)_L$ & $\bm{2}$ & $\bm{1}$ & $\bm{1}$ & $\bm{2}$ & $\bm{1}$ & $\bm{2}$ & $\bm{1}$ & $\bm{2}$ & $\bm{2}$ & $\bm{1}$  & $\bm{1}$   \\
    \hline
    $U(1)_Y$ & $\frac16$ & $\frac23$ & $-\frac13$ & $-\frac12$ & $-1$ & $-\frac12$ & $0$ & $\frac12$ & $\frac12$ & $0$  & $0$  \\
    \hline
    $U(1)_H$ & $0$ & $0$ & $0$ & $0$ & $0$ & $-x$ & $ \frac12 x$ & $0$ & $\frac32 x$ & $x$ & $\frac12 x$ \\
    \hline
  \end{tabular}
  \caption{Field contents and charge assignment in the model.   \label{tab:table1}}
\end{table}

In this model we consider hidden $U(1)_H$ gauge symmetry in addition to the SM gauge symmetry.
Here we introduce the vector like extra lepton doublet $L' = (N', E')$ and SM singlet fermions $N_{L,R}$ which have $U(1)_H$ charge $x$ and $x/2$ respectively.
In scalar sector, $SU(2)_L$ doublet $\eta$, singlet $\varphi$ and singlet $\chi$ are introduced whose $U(1)_H$ charges are $x/2$, $x$ and $x/2$.  
The field contents and their charge assignments are summarized in Table.~\ref{tab:table1}.
In this model, $\varphi$ and the SM Higgs field $\Phi$ develop VEVs spontaneously breaking $U(1)_H$ and electroweak symmetry.
The scalar fields are written by 
\begin{align}
& \Phi = \begin{pmatrix} G^+ \\ \frac{1}{\sqrt{2}} (v + \tilde h + i G_Z) \end{pmatrix}, \quad \eta = \begin{pmatrix} \eta^+ \\ \frac{1}{\sqrt{2}} ( \eta_R + i \eta_I) \end{pmatrix} \nonumber \\
& \varphi =  \frac{1}{\sqrt{2}} (v_\varphi + \varphi_R + i G_{Z^\prime}), \quad  \chi = \frac{1}{\sqrt{2}} ( \chi_R + i \chi_I),
\label{eq:scalars}
\end{align}
where $v$ and $v_{\varphi}$ are VEVs of $\Phi$ and $\varphi$, and $G^\pm$, $G_{Z}$ and $G_{Z^\prime}$ are Nambu-Goldstone(NG) bosons absorbed by $W^\pm$, $Z$ and $Z^\prime$ gauge bosons.
Note that $Z_2$ symmetry remains after $U(1)_H$ symmetry breaking where $N_{L/R}$, $\eta$ and $\chi$ are parity odd under the $Z_2$ while the other particles are parity even.
Thus the lightest $Z_2$ odd particle can be our DM candidate.

In this model, Yukawa interactions for leptons~\footnote{Yukawa interactions for quarks are the same as the SM one and we do not discuss in the paper.} and scalar potential are given by
\begin{align}
\mathcal{L} \supset \ & y_\ell \bar L_L \Phi e_R + M_{L'} \bar L'_L L'_R + M_N \bar N_R N_L + y_D \bar L_L L'_R \varphi + y_{R} \bar L'_L \tilde \eta N_R + y_L \bar L'_R \tilde \eta N_L \nonumber \\
& + y_{\varphi_R} \bar N^c_R N_R \varphi^* + y_{\varphi_L} \bar N^c_L N_L \varphi^*  + h.c. \, , \\
V = \ & m_\Phi^2 \Phi^\dagger \Phi + m_\eta^2 \eta^\dagger \eta + m_\varphi^2 \varphi^* \varphi  + m_\chi^2 \chi^* \chi  + \mu(\chi^2 \varphi^* + h.c.) \nonumber \\
& + \lambda_0((\Phi^\dagger \eta) \chi^* \varphi^* + h.c.) + \lambda_\Phi (\Phi^\dagger \Phi)^2 +  \lambda_\eta (\eta^\dagger \eta)^2 + \lambda_\chi (\chi^* \chi)^2  + \lambda_\varphi (\varphi^* \varphi)^2 \nonumber\\
& + \lambda_{\Phi \eta} (\Phi^\dagger \Phi)(\eta^\dagger \eta) + \tilde \lambda_{\Phi \eta} (\Phi^\dagger \eta)(\eta^\dagger \Phi) + \lambda_{\Phi \varphi} (\Phi^\dagger \Phi)(\varphi^* \varphi) + \lambda_{\Phi\chi} (\Phi^\dagger \Phi)(\chi^* \chi) \nonumber\\
& + \lambda_{\eta \varphi }(\eta^\dagger \eta)(\varphi^* \varphi) + \lambda_{\eta \chi} (\eta^\dagger \eta)(\chi^* \chi) + \lambda_{\varphi \chi} (\varphi^* \varphi)(\chi^* \chi) \, . 
\label{eq:potential}
\end{align}

\subsection{Scalar sector}

Here we consider scalar sector in the model and formulate mass spectrum and corresponding mass eigenvalues.
Firstly the VEVs of $\Phi$ and $\varphi$ are calculated from scalar potential Eq.~(\ref{eq:potential}) by solving the condition $\partial V/\partial v = \partial V/\partial v_\varphi = 0$.
Then we obtain the conditions 
\begin{align}
& m_\Phi^2 +  \lambda_\Phi v^2 + \frac{1}{2} \lambda_{\Phi \varphi} v_\varphi^2 = 0, \\ 
& m_\varphi^2 +  \lambda_\varphi v_\varphi^2 + \frac{1}{2} \lambda_{\Phi \varphi} v^2 = 0.  
\end{align}
Note that the conditions $m_\Phi^2, m_\varphi^2 < 0$ are required to make square of VEVs positive definite.
In addition we assume $m_\chi^2, m_{\eta}^2 >0$ so that $\chi$ and $\eta$ do not develop non-zero VEVs.

After spontaneous symmetry breaking, we obtain mass matrix for CP-even neutral scalar fields with $Z_2$-even parity such that
\begin{equation}
\mathcal{L} \supset \frac12 \begin{pmatrix} \tilde h \\ \varphi_R \end{pmatrix}^T 
\begin{pmatrix} 2 \lambda_\Phi v^2 & \lambda_{\Phi \varphi} v v_\varphi \\ \lambda_{\Phi \varphi} v v_\varphi & 2 \lambda_\varphi v_\varphi^2 \end{pmatrix}.
\begin{pmatrix} \tilde h \\ \varphi_R \end{pmatrix}
\end{equation} 
Diagonalizing the mass matrix, the mass eigenstates are given by
\begin{align}
m_{h,H}^2 = \lambda_\Phi v^2 + \lambda_\varphi v_\varphi^2 \pm \sqrt{(\lambda_\Phi v^2 - \lambda_\varphi v_\varphi^2)^2 + \lambda_{\Phi \varphi}^2 v^2 v_\varphi^2 },
\end{align}
where we identify $m_h = 125$ GeV as the SM Higgs mass.
The mass eigenstates and mixing are written as 
\begin{align}
& \begin{pmatrix} \tilde h \\ \varphi_R \end{pmatrix} = \begin{pmatrix} \cos \beta & - \sin \beta \\ \sin \beta & \cos \beta \end{pmatrix} \begin{pmatrix} h \\ H \end{pmatrix}, \nonumber \\
& \tan 2\beta = \frac{v v_{\varphi}\lambda_{\Phi\varphi}}{v^2 \lambda_\Phi - v^2_{\varphi}\lambda_\varphi},
\end{align}
where $h$ is identified as the SM Higgs boson.

The mass for $Z_2$-odd charged scalar field is given by
\begin{align}
m_{\eta^\pm}^2 = m_\eta^2 + \frac{1}{2} \lambda_{\Phi \eta}   v^2 + \frac{1}{2} \lambda_{\eta \varphi} v_\varphi^2,
\end{align}
where the corresponding mass eigenstate is $\eta^\pm$ given in Eq.~(\ref{eq:scalars}).
We also obtain mass matrix for $Z_2$-odd CP-odd scalar fields such as 
\begin{align}
& \mathcal{L} \supset \frac12 \begin{pmatrix} \eta_I \\ \chi_I \end{pmatrix}^T 
\begin{pmatrix} m_{\eta^\pm}^2 + \frac{1}{2} \tilde \lambda_{\Phi \eta} v^2 & -\frac12 \lambda_{0} v v_\varphi \\  -\frac12 \lambda_{0} v v_\varphi & m_{\chi'}^2 \end{pmatrix}
\begin{pmatrix} \eta_I \\ \chi_I \end{pmatrix}, \nonumber \\
& m^2_{\chi'} \equiv m_\chi^2 + \frac12 \lambda_{\Phi\chi} v^2 + \frac12 \lambda_{\varphi \chi} v_\varphi^2 - \sqrt2 \mu v_\varphi.
\end{align}
Diagonalizing the mass matrix, the mass eigenstates are given by
\begin{eqnarray}
m_{A_1,A_2}^2 &=& \frac{1}{4}\Big[ 2m_{\eta^\pm}^2 + 2 m_{\chi^\prime}^2 + v^2 \tilde{\lambda}_{\Phi\eta} \\ \nonumber
&\pm & \sqrt{(2m_{\eta^\pm}^2 + 2m_{\chi^\prime}^2 + v^2 \tilde{\lambda}_{\Phi\eta})^2 -4(4m_{\eta^\pm}^2 m_{\chi^\prime}^2 -v^2v_{\varphi}^2 \lambda_0^2 + 2m_{\chi^\prime}^2 v^2 \tilde{\lambda}_{\Phi\eta})} \Big],
\end{eqnarray}
The mass eigenstates and mixing are written as 
\begin{align}
& \begin{pmatrix} \eta_I \\ \chi_I \end{pmatrix} = \begin{pmatrix} \cos \gamma & - \sin \gamma \\ \sin \gamma & \cos \gamma \end{pmatrix} \begin{pmatrix} A_1 \\ A_2 \end{pmatrix},  \nonumber  \\
& \tan 2 \gamma = \frac{v v_{\varphi}\lambda_0}{m^2_{\chi^\prime} - m^2_{\eta^\pm} - \frac{1}{2}v^2 \tilde{\lambda}_{\Phi\eta}}.
\end{align}
Furthermore mass matrix for $Z_2$-odd CP-even scalar fields are obtained  as follows 
\begin{align}
& \mathcal{L} \supset \frac12 \begin{pmatrix} \eta_R \\ \chi_R \end{pmatrix}^T 
\begin{pmatrix} m_{\eta^\pm}^2 + \frac{1}{2} \tilde \lambda_{\Phi \eta} v^2 & \frac12 \lambda_{0} v v_\varphi \\  \frac12 \lambda_{0} v v_\varphi & m_{\chi''}^2 \end{pmatrix}
\begin{pmatrix} \eta_R \\ \chi_R \end{pmatrix},  \nonumber \\
& m^2_{\chi''} \equiv m_\chi^2 + \frac12 \lambda_{\Phi\chi} v^2 + \frac12 \lambda_{\varphi \chi} v_\varphi^2 + \sqrt2 \mu v_\varphi.
\end{align}
Diagonalizing the mass matrix, the mass eigenstates are given by
\begin{eqnarray}
m_{H_1,H_2}^2 &=& \frac{1}{4}\Big[ 2 m_{\eta^\pm}^2 + 2 m_{\chi^{\prime\prime}}^2 + v^2 \tilde{\lambda}_{\Phi\eta} \\ \nonumber
 &\pm& \sqrt{(2m_{\eta^\pm}^2 + 2m_{\chi^{\prime\prime}}^2 + v^2 \tilde{\lambda}_{\Phi\eta})^2 -4(4m_{\eta^\pm}^2 m_{\chi^{\prime\prime}}^2 - v^2 v_{\varphi}^2 \lambda_0^2 + 2m_{\chi^{\prime\prime}}^2v^2\tilde{\lambda}_{\Phi\eta})} \Big],
\end{eqnarray}
The mass eigenstates and mixing are written as 
\begin{align}
& \begin{pmatrix} \eta_R \\ \chi_R \end{pmatrix} = \begin{pmatrix} \cos \alpha & - \sin \alpha \\ \sin \alpha & \cos \alpha \end{pmatrix} \begin{pmatrix} H_1 \\ H_2 \end{pmatrix}, \\
& \tan 2 \alpha = \frac{v v_\varphi \lambda_0}{m^2_{\eta^\pm} - m^2_{\chi^{\prime\prime}} + \frac{1}{2}v^2 \tilde{\lambda}_{\Phi\eta}}.
\end{align}
We find that the mixing $\sin \alpha$ and $\sin \gamma$ can be written in terms of mass eigenvalues such that
\begin{eqnarray}
\cos^2\alpha &=& \frac{(m^2_{A_2} - m^2_{H_2})(m^2_{H_2} - m^2_{A_1})}{(m^2_{A_2} + m^2_{A_1} - m^2_{H_2} - m^2_{H_1})(m^2_{H_2} - m^2_{H_1})} \nonumber\\
\cos^2 \gamma &=& \frac{(m^2_{A_2} - m^2_{H_2})(m_{A_2}^2 - m^2_{H_1})}{(m^2_{A_2} + m^2_{A_1} - m^2_{H_2} - m^2_{H_1})(m^2_{A_2} -m^2_{A_1})}.
\end{eqnarray}
Note that we can write some parameters in the scalar potential by VEVs, scalar masses and mixings as summarized in Appendix where
the independent parameters are $\{m_h, m_H, m_{H_1}, m_{H_2}, m_{A_1}, m_{A_2}, m_{\eta^\pm}, v, v_{\varphi}, \cos\beta$, $\lambda_{\Phi \chi},\lambda_{\varphi\chi},\lambda_{\Phi\eta}, \lambda_{\eta\varphi}, \lambda_{\eta\chi}, \lambda_{\eta}, \lambda_{\chi} \}$.

\subsection{Gauge sector}

  The most general $U(1)$ gauge Lagrangian including the kinetic mixing term is 
\begin{equation}
\mathcal{L}_{\text{gauge}}= -\frac{1}{4}B_{\mu\nu}B^{\mu\nu} - \frac{1}{4}B^\prime_{\mu\nu}B^{\prime\mu\nu} - \frac{1}{2}\epsilon B_{\mu\nu}B^{\prime \mu\nu} 
\label{Lgauge}
\end{equation} 
where $B_{\mu\nu}$ and $B^\prime_{\mu\nu}$ are the field strength tensors of $U(1)_Y$ and $U(1)_H$ gauge symmetries.  We can diagonalize Eq.~\eqref{Lgauge} by the following transformation

\begin{eqnarray}
\left(\begin{array}{c}
\tilde{B}^\prime_\mu\\
\tilde{B}_\mu\\
\end{array}\right)=\left(\begin{array}{cc}
\sqrt{1 - \epsilon^2} & 0 \\
\epsilon & 1 \\
\end{array}\right)\left(\begin{array}{c}
B^\prime_\mu\\
B_\mu\\
\end{array}\right)
\label{Gl2R}
\end{eqnarray}
where $\epsilon$ is a dimensionless quantity ($\epsilon << 1$) and we parameterize $\rho=-\frac{\epsilon}{\sqrt{1-\epsilon^2}}$. Under the transformation Eq.~\eqref{Gl2R}, the gauge Lagrangian can be written as 
\begin{eqnarray}
\mathcal{L}_{\text{gauge}}=-\frac{1}{4}\tilde{B}_{\mu\nu}\tilde{B}^{\mu\nu}-\frac{1}{4}\tilde{B}^\prime_{\mu\nu}\tilde{B}^{\prime\mu\nu}
\end{eqnarray}   
where $\tilde{B}_{\mu\nu}=\partial_\mu \tilde{B}_\nu - \partial_\nu \tilde{B}_\mu$ and $\tilde{B}^\prime_{\mu\nu}=\partial_\mu \tilde{B}^\prime_\nu - \partial_\nu \tilde{B}^\prime_\mu$.

The kinetic term of the scalar fields are 
\begin{eqnarray}
\mathcal{L}_{\text{kin}} = (D_{\mu}\Phi)^\dagger (D^{\mu} \Phi) + (D_{\mu}\eta)^\dagger (D^{\mu} \eta) + (D_{\mu}\varphi)^\dagger (D^{\mu} \varphi) + (D_{\mu}\chi)^\dagger (D^{\mu} \chi).
\label{eq:scalar-kinetic}
\end{eqnarray}
The covariant derivatives of scalar fields are written by
\begin{eqnarray}
D_{\mu}\Phi &=& \Big(\partial_\mu + ig\frac{\tau^a}{2}W_{\mu}^a + i\frac{g^\prime}{2}\tilde{B}_\mu + i\frac{g^\prime}{2}\rho \tilde{B}^\prime_\mu  \Big)\Phi, \nonumber \\
D_{\mu}\eta &=& \Big( \partial_\mu + i g \frac{\tau^a}{2} W_{\mu}^a +  i\frac{g^\prime}{2}\tilde{B}_\mu + \Big( i \frac{g^\prime}{2} \rho -i g^{\prime\prime} \frac{3x}{2} \frac{\rho}{\epsilon} \Big)\tilde{B}^\prime_\mu \Big)\eta, \nonumber \\
D_{\mu}\varphi &=& \Big( \partial_\mu -ig^{\prime\prime}x \frac{\rho}{\epsilon}\tilde{B}^\prime_\mu \Big)\varphi, \nonumber \\
D_\mu \chi &=& \Big( \partial_\mu - i \frac{g^{\prime\prime}}{2}x\frac{\rho}{\epsilon}\tilde{B}^\prime_mu \Big) \chi,
\label{Covariant derivatives}
\end{eqnarray}
where $g$, $g'$ and $g''$ are gauge couplings of $SU(2)_L$, $U(1)_Y$ and $U(1)_H$, $W^a$ is the $SU(2)_L$ gauge field, and $\tau^a$ is the Pauli matrix. 

The scalar fields $H$ and $\varphi$ get VEVs whereas other fields do not develop VEVs to preserve the $Z_2$ symmetry. The masses of the gauge bosons come from the first and third term of Eq.~\eqref{eq:scalar-kinetic}. The mass term of the neutral gauge bosons written in the basis of neutral gauge fields $(W_\mu^3, \tilde{B}_\mu, \tilde{B}^\prime_\mu)$ is 
\begin{eqnarray}
\mathcal{L}_{\text{gauge}}^{\text{mass}} = \frac{1}{2}
\left(\begin{array}{c}
W_\mu^3 \\
\tilde{B}_\mu\\
\tilde{B}^\prime_\mu
\end{array}\right)^T M^2_{\text{gauge}}
\left(\begin{array}{c}
W_\mu^3 \\
\tilde{B}_\mu\\
\tilde{B}^\prime_\mu
\end{array}\right)
\end{eqnarray} 
where
\begin{eqnarray}
M^2_{\text{gauge}}=\frac{1}{4}
\left(\begin{array}{ccc}
g^2 v^2 & -gg^\prime v^2 & -gg^\prime v^2 \rho \\
-g g^\prime v^2 & g^{\prime 2} v^2 & g^{\prime 2} v^2 \rho\\
-g g^\prime v^2 \rho & g^{\prime 2} v^2 \rho & g^{\prime 2} v^2 \rho^2 + 4 g^{\prime\prime 2}v^2_{\varphi}x^2\frac{\rho^2}{\epsilon^2} \\
\end{array}\right).
\label{masss matrix gauge boson}
\end{eqnarray}
Here we parameterize 
\begin{eqnarray}
M_{Z^\prime}^2 = \frac{1}{4} g^{\prime 2} v^2 \rho^2 + g^{\prime\prime 2} v_{\varphi}^2 x^2 \frac{\rho^2}{\epsilon^2}. 
\end{eqnarray}
We rotate the fields $(W_\mu^3,\tilde{B}_\mu)$ by Weinberg angle $\theta_W$ to obtain the massless photon field $A_\mu$
\begin{eqnarray}
\left(\begin{array}{c}
W^3_\mu \\
\tilde{B}_\mu \\
\end{array}\right) =
\left(\begin{array}{cc}
\cos\theta_W & \sin\theta_W \\
-\sin\theta_W & \cos\theta_W\\
\end{array}\right)
\left(\begin{array}{c}
\tilde{Z}_\mu \\
A_\mu \\
\end{array}\right).
\end{eqnarray}
and the mass matrix for the massive neutral gauge bosons
\begin{eqnarray}
&& \mathcal{L}^{\text{mass}}_{\text{gauge}} = \frac{1}{2}
\left(\begin{array}{c}
\tilde{Z}_\mu \\
\tilde{B}^\prime_\mu \\
\end{array}\right)^T
\left(\begin{array}{cc}
M_{Z,SM}^2 & -\Delta^2 \\
-\Delta^2 & M_{Z'}^2
\end{array}\right)
\left(\begin{array}{c}
\tilde{Z}_\mu \\
\tilde{B}^\prime_\mu \\
\end{array}\right)
\label{masss matrix gauge boson 2}
\end{eqnarray}
where $\Delta^2 = \frac{1}{4} g^\prime v^2 \rho \sqrt{g^2 +g^{\prime 2}}$ 
and $M_{Z,SM}^2 = \frac{1}{4} v^2 (g^2 + g^{\prime 2})$.
The physical masses of the neutral gauge bosons are 
\begin{eqnarray}
m_Z^2 &=& \frac{1}{2}\Big[M_{Z,SM}^2 + M_Z^{\prime 2} + \sqrt{(M_{Z,SM}^2 - M_{Z^\prime}^2)^2 + 4\Delta^4} \Big] \nonumber \\
m_{Z^{\prime}}^2 &=& \frac{1}{2}\Big[M_{Z,SM}^2 + M_Z^{\prime 2} - \sqrt{(M_{Z,SM}^2 - M_{Z^\prime}^2)^2 + 4\Delta^4} \Big].
\end{eqnarray}
In the limit $\epsilon \to 0$, we have $m_{Z} \approx M_{Z,SM}$ and $m_{Z^\prime} \approx M_{Z^\prime}$.
The mass matrix in Eq.~\eqref{masss matrix gauge boson 2} can be diagonalized by rotation matrix 
\begin{eqnarray}
\left(\begin{array}{c}
\tilde{Z}_\mu\\
\tilde{B}^\prime_\mu\\
\end{array}\right)&=&
\left(\begin{array}{cc}
\cos\theta & \sin\theta\\
-\sin\theta & \cos\theta \\
\end{array}\right)
\left(\begin{array}{c}
Z_\mu\\
Z^\prime_\mu\\
\end{array}\right) \\
\tan 2\theta &=& \frac{2\Delta^2}{M_{Z,SM}^2 - M_{Z^\prime}^2}
\end{eqnarray}
where $Z_\mu$ and $Z^\prime_\mu$ are the two physical gauge bosons correspond to the SM $Z$ boson and extra gauge boson. The $Z-Z^\prime$ mixing would disappear in the limit $\epsilon \rightarrow 0$. In summary, the original unphysical gauge fields are transformed into mass eigenstates by
\begin{eqnarray}
\left(\begin{array}{c}
W_\mu^3 \\
B_\mu \\
B^\prime _\mu \\
\end{array}\right) &=&
\left(\begin{array}{ccc}
R^w_{11} & R^w_{12} & R^w_{13}\\
R^w_{21} + R^w_{31}\rho & R^w_{22} + R^w_{32}\rho & R^w_{23} + R^w_{33}\rho \\
-R^w_{31}\frac{\rho}{\epsilon} & -R^w_{32}\frac{\rho}{\epsilon} & -R^w_{33}\frac{\rho}{\epsilon}\\ 
\end{array}\right)
\left(\begin{array}{c}
Z_\mu \\
A_\mu \\
Z^\prime_\mu \\
\end{array} \right) 
\label{eq:gauge field}
\\
 R^w_{ab}&=&\begin{pmatrix}
 \cos{\theta_w}\cos{\theta} & \sin{\theta_w} & \cos{\theta_w}\sin{\theta} \\
-\sin{\theta_w}\cos{\theta} & \cos{\theta_w} & -\sin{\theta_w}\sin{\theta} \\
-\sin{\theta}& 0 &\cos{\theta}
\end{pmatrix}_{ab},
\label{Rw matrix}
\end{eqnarray}
where $a(b) = 1,2,3$.

\subsection{Constraints in boson sector}
{
Here we consider constraints in boson sector such as perturbativity, unitarity, vacuum stability, $T$-parameter and Higgs invisible decay.
We mainly focus on two Higgs doublet $\{\Phi, \eta \}$ and $\varphi$ since parameters associated with $\chi$ are completely free and phenomenological constraints can be easily satisfied. 
The constraints from unitarity and perturbativity are given by~\cite{Bian:2017xzg,Muhlleitner:2016mzt} 
\begin{align}
& |\lambda_{\Phi,\eta,\varphi, \Phi\eta }| \leq 4 \pi, \quad |\lambda_{\Phi\varphi, \Phi\eta}| \leq 8 \pi, \quad |\lambda_{\Phi\eta} \pm \tilde \lambda_{\Phi \eta}| \leq 8 \pi, \quad  |\lambda_{\Phi \eta} + 2 \tilde \lambda_{\Phi\eta} | \leq 8 \pi, \nonumber \\
& \sqrt{|\lambda_{\Phi \eta} (\lambda_{\Phi \eta} + 2 \tilde \lambda_{\Phi \eta})|} \leq 8 \pi,  \quad
\left| \lambda_\Phi + \lambda_\eta \pm \sqrt{(\lambda_\Phi - \lambda_\eta)^2 + \tilde \lambda^2_{\Phi\eta}} \right| \leq 8 \pi, \nonumber \\
& \left| \lambda_\Phi + \lambda_\eta \pm \sqrt{(\lambda_\Phi - \lambda_\eta)^2 } \right| \leq 8 \pi, \quad a_{1,2,3} \leq 8 \pi,
\end{align}
where $a_{1,2,3}$ are obtained as the solutions of the following equation
\begin{align}
& x^3 - 2x^2 (3 \lambda_\Phi + 3 \lambda_\eta + 2 \lambda_{\varphi})  \nonumber \\
& -  x (2 \lambda_{\Phi \varphi}^2 + 2 \lambda_{\eta \varphi}^2 - 36 \lambda_\Phi \lambda_\eta - 24 \lambda_\Phi \lambda_{\Phi \varphi} - 24 \lambda_1 \lambda_{\Phi \varphi} + 4 \lambda_{\Phi\eta}^2 + 4\lambda_{\Phi \eta} \tilde \lambda_{\Phi \eta} + \tilde \lambda_{\Phi \eta}^2 ) \nonumber \\
& + 4 (3 \lambda_{\Phi \varphi}^2 \lambda_\eta - \lambda_{\Phi \varphi} \lambda_{\eta \varphi} (2 \lambda_{\Phi \eta} + \tilde \lambda_{\Phi \eta}) + 3 \lambda_{\eta \varphi}^2 \lambda_\Phi + \lambda_{\varphi} ((2 \lambda_{\Phi \eta} + \tilde \lambda_{\Phi \eta})^2 - 36 \lambda_\Phi \lambda_\eta)) =0.
\end{align}
We can also writhe the conditions to guarantee vacuum stability such that~\cite{Muhlleitner:2016mzt}:
\begin{align}
& \Omega_1 \cup \Omega_2 \\
& \Omega_1 = \biggl\{ \lambda_{\Phi,\eta,\varphi} >0, \ 2 \sqrt{\lambda_\Phi \lambda_{\varphi} } + \lambda_{\Phi\varphi} >0, \ 2 \sqrt{\lambda_\eta \lambda_{\varphi} } + \lambda_{\eta \varphi} >0,   \nonumber \\
& \qquad \qquad 2\sqrt{\lambda_\Phi \lambda_\eta} + \lambda_{\Phi \eta} > 0, \  \lambda_{\Phi\varphi} + \sqrt{\frac{\lambda_\Phi}{\lambda_\eta}} \lambda_{\Phi\varphi}  \geq 0 \biggr\} \\
& \Omega_2 = \biggl\{ \lambda_{\Phi,\eta,\varphi} >0, \ 2 \sqrt{\lambda_\Phi \lambda_{\varphi} } \geq \lambda_{\eta \varphi}  > -2 \sqrt{\lambda_\eta \lambda_{\varphi} }, 
\ 2 \sqrt{\lambda_\Phi \lambda_{\varphi} } > -\lambda_{\Phi \varphi}  \geq \sqrt{\frac{\lambda_\Phi}{\lambda_\eta}} \lambda_{\eta \varphi},   \nonumber \\
& \qquad \qquad \sqrt{(\lambda_{\Phi \varphi}^2 - 4 \lambda_\Phi \lambda_{\varphi})(\lambda_{\eta \varphi}^2 - 4 \lambda_\eta \lambda_{\varphi}) } > \lambda_{\Phi \varphi} \lambda_{\eta \varphi} - 2 \lambda_{\Phi \eta} \lambda_{\varphi} \biggr\}.
\label{vac}
\end{align}
In addition we impose inert condition for $\eta$ requiring the potential be bounded below~\cite{Nomura:2019yft}
\begin{equation}
-\frac{2}{\sqrt{21}}\sqrt{\lambda_\Phi\lambda_\eta} < \lambda_{\Phi\eta}+ \tilde \lambda_{\Phi\eta},\quad 
-\sqrt6 \lambda'_{\Phi\eta}<|\lambda_{\Phi\eta}+ \tilde \lambda_{\Phi\eta}|,
\quad -\frac{2}{\sqrt{21}}\sqrt{\lambda_\eta\lambda_\varphi} < \lambda_{\eta\varphi}.
\end{equation}
The contribution to $T$-parameter from scalar loops can be written as~\cite{Barbieri:2006dq}
\begin{align}
& \Delta T = \frac{1}{32 \alpha v^2} \left( G[m_{\eta^\pm}^2, m_{A_1}^2] \cos^2 \gamma + G[m_{\eta^\pm}^2, m_{A_2}^2] \sin^2 \gamma \right.  \nonumber \\
& \qquad \qquad \qquad \left. + G[m_{\eta^\pm}^2, m_{H_1}^2] \cos^2 \alpha + G[m_{\eta^\pm}^2, m_{H_2}^2] \sin^2 \alpha \right), \\
& G[m_1^2,m_2^2] \equiv \frac{m_1^2 +m_2^2}{2} - \frac{m_1^2 m_2^2}{m_1^2 - m_2^2} \ln \left[ \frac{m_1^2}{m_2^2} \right],
\end{align}
where $\alpha$ is the fine structure constant. Here we adopt the constraint~\cite{Tanabashi:2018oca} 
\begin{equation}
-0.01 < \Delta T < 0.11,
\end{equation}
where vanishing $U$-parameter is assumed.

In our analysis	below, we consider relatively small $v_\varphi (\ll v)$ to satisfy constraints from $L'$--$L$ mixing induced by $\bar L_L L'_R \varphi$ Yukawa interaction.
Thus $Z'$ and $H$ masses tend to be light and we need to consider decay processes of $h \to Z'Z'$ and $h \to H H$.
The decay widths are given by
\begin{align}
& \Gamma(h \to Z' Z') = \frac{1}{8 \pi} \frac{m_{Z'}^4}{v_\varphi^2 m_h} s^2_\beta
\left( 2 + \frac{m_h^4}{4 m_{Z'}^4} \left( 1- \frac{2 m_{Z'}^2}{m_h^2} \right)^2 \right) \sqrt{1 - \frac{4 m_{Z'}^2}{m_h^2}}, \\
& \Gamma(h \to HH) = \frac{C_{hHH}^2}{16 \pi m_h} \sqrt{1 - \frac{4 m_{H}^2}{m_h^2}}, \\
& C_{hHH} \equiv \frac{\lambda_{\Phi \varphi}}{2} \left( v (c^2_\beta - 2 c_\beta s^2_\beta ) + v_\varphi (s^3_\beta - 2  c^2_\beta s_\beta)  \right) + \frac92 s_\beta c_\beta (\lambda_\Phi s_\beta v + \lambda_\varphi c_\beta v_\varphi),
\end{align}
where $c_\beta= \cos \beta(s_\beta = \sin \beta)$.
Here $H$ mainly decays into $Z'Z'$ mode and $Z'$ decays into the SM particles via kinetic mixing or Lepton mixing effect.
In our scenario we consider kinetic mixing is tiny and $Z'$ is long lived to escape detector at the LHC experiments; $Z'$ could decay into leptons via $L'$--$L$ mixing discussed below and it is also considered to be tiny.
Then we apply constrains from invisible decay of $h$ for these decay branching ratio (BR) where the upper bound is taken to be $BR(h \to inv) <0.19$~\cite{Aaboud:2019rtt,Sirunyan:2018owy}.
We obtain BRs of $h \to Z'Z'$ and $h \to HH$ modes in the case of $m_{Z' H}^2 \ll m_{h}^2$ as follows 
\begin{align}
BR(h \to Z'Z') \simeq 4.6 \times 10^{-4} \left( \frac{s_\beta}{10^{-5}} \right)^2 \left( \frac{1 \ {\rm GeV}}{v_\varphi} \right)^2, \\
BR(h \to HH) \simeq 9.3 \times 10^{-4} \left( \frac{s_\beta}{10^{-5}} \right)^2 \left( \frac{1 \ {\rm GeV}}{v_\varphi} \right)^2, 
\end{align}
where we used the SM Higgs decay width $\Gamma_h \simeq 4.2$ MeV.
Thus very small $s_\beta$ is required to avoid constraints from the SM Higgs decay.

Finally we discuss allowed region from the above constraints where we assume $v_\varphi \in [1, 5]$ GeV, $g'' \in [0.001,1]$, $m_{H} \in [0.1, 50]$ GeV, $m_{H_1} \in [150, 1500]$ GeV, $m_{H_2} \in [0.1, 1500]$ GeV,
$m_{A_1} \in [150, 1500]$ GeV and $m_{A_2} \in [0.1, 1500]$ GeV. We also assume small mixing among inert scalar bosons as $\sin \alpha < 0.1$ and $\sin \gamma < 0.1$.
We find that mass of $H$ is light as $m_H \lesssim 20$ GeV due to small $v_\varphi$ and unitarity condition as shown in left plot of Fig.~\ref{fig:const1}.
Also $\Delta T$ constraint requires $|m_{H_1} - m_{A_1}| \lesssim 4$ GeV and $|m_{\eta^{\pm}} - m_{H_1}| \lesssim 120$ GeV.
The mixing between the SM Higgs and $H$ should be suppressed to avoid constraint from invisible decay as shown in right plot  of Fig.~\ref{fig:const1}.
The other scalar mass values are not constrained.

\begin{figure}[tb]
\includegraphics[width=7cm]{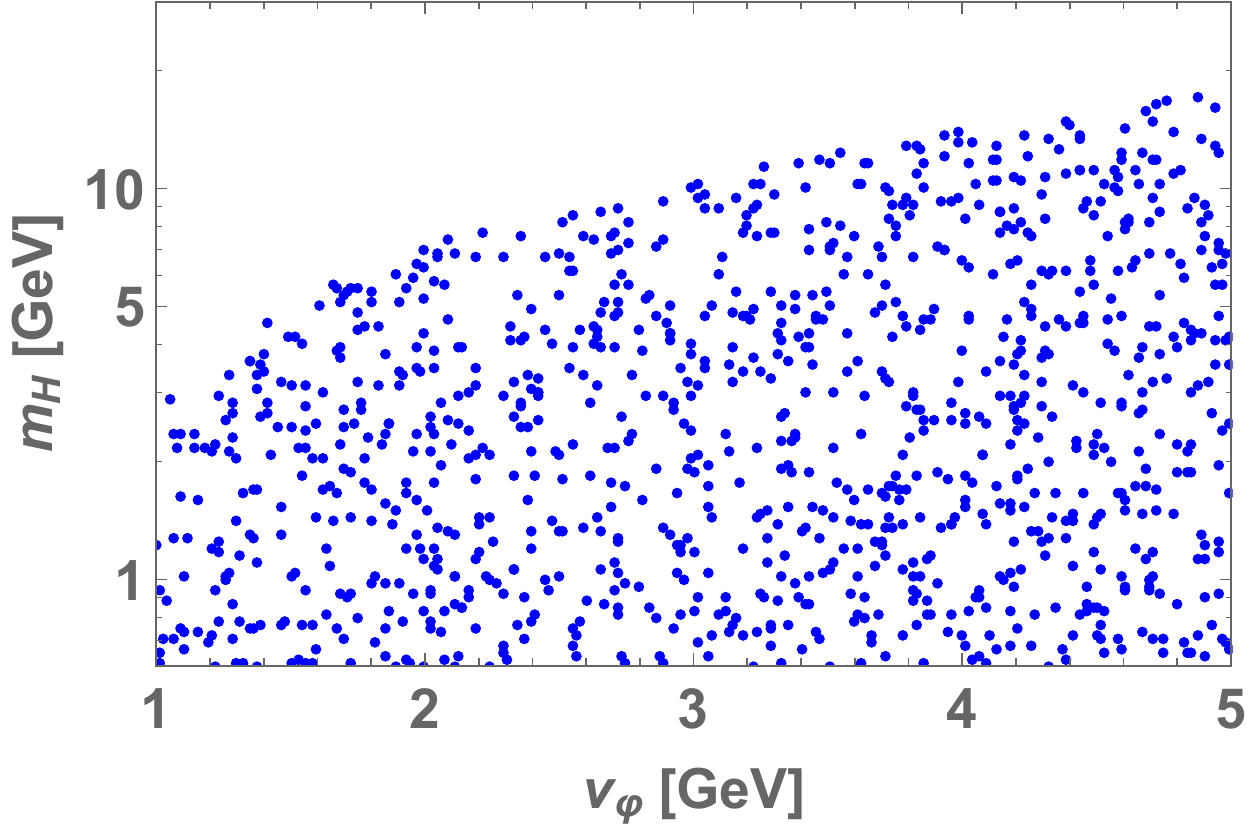}
\includegraphics[width=8cm]{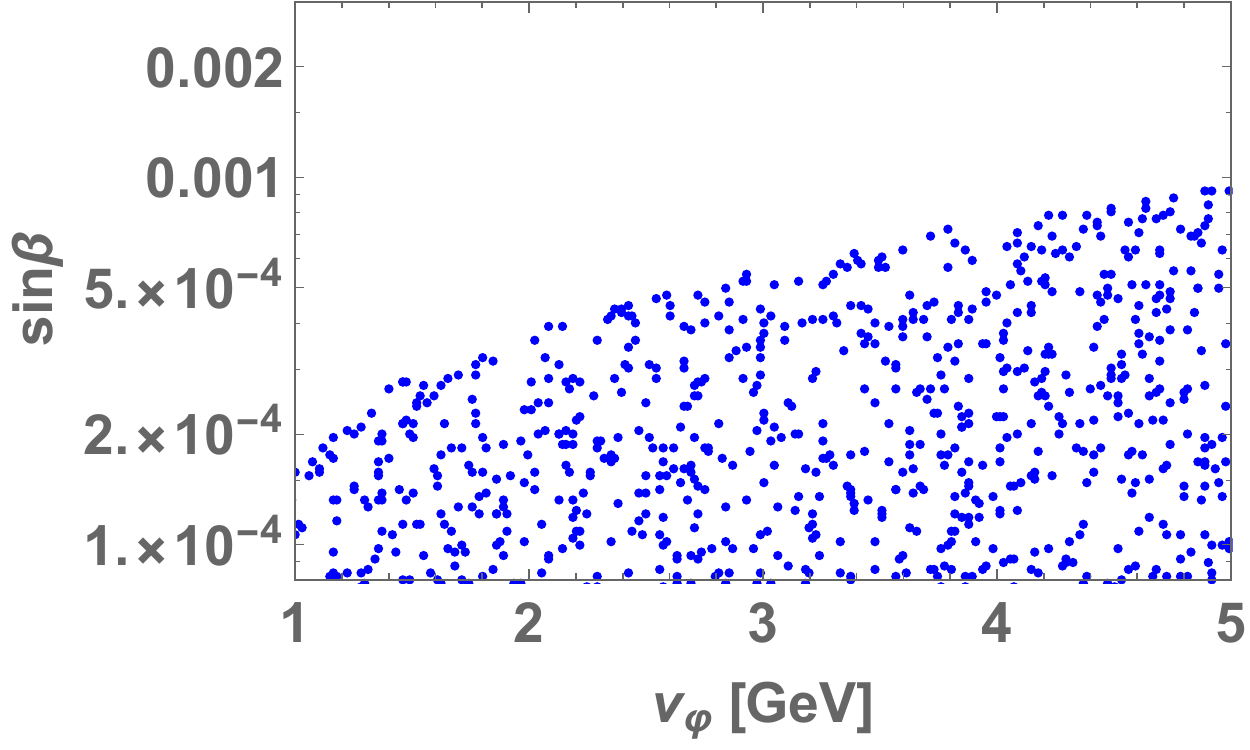}
\caption{Allowed region on $\{v_\varphi, m_H\}$ and $\{v_\varphi, \sin \beta\}$ planes.}
\label{fig:const1}
\end{figure}

}

\section{Neutrino mass and lepton flavor violation}

In this section we formulate neutrino mass matrix, LFV processes and muon $g-2$.
We then carry out numerical analysis to search for allowed parameter ranges taking into account experimental constraints.

\subsection{Neutrino mass generation}

\begin{figure}[tb]
\includegraphics[width=80mm]{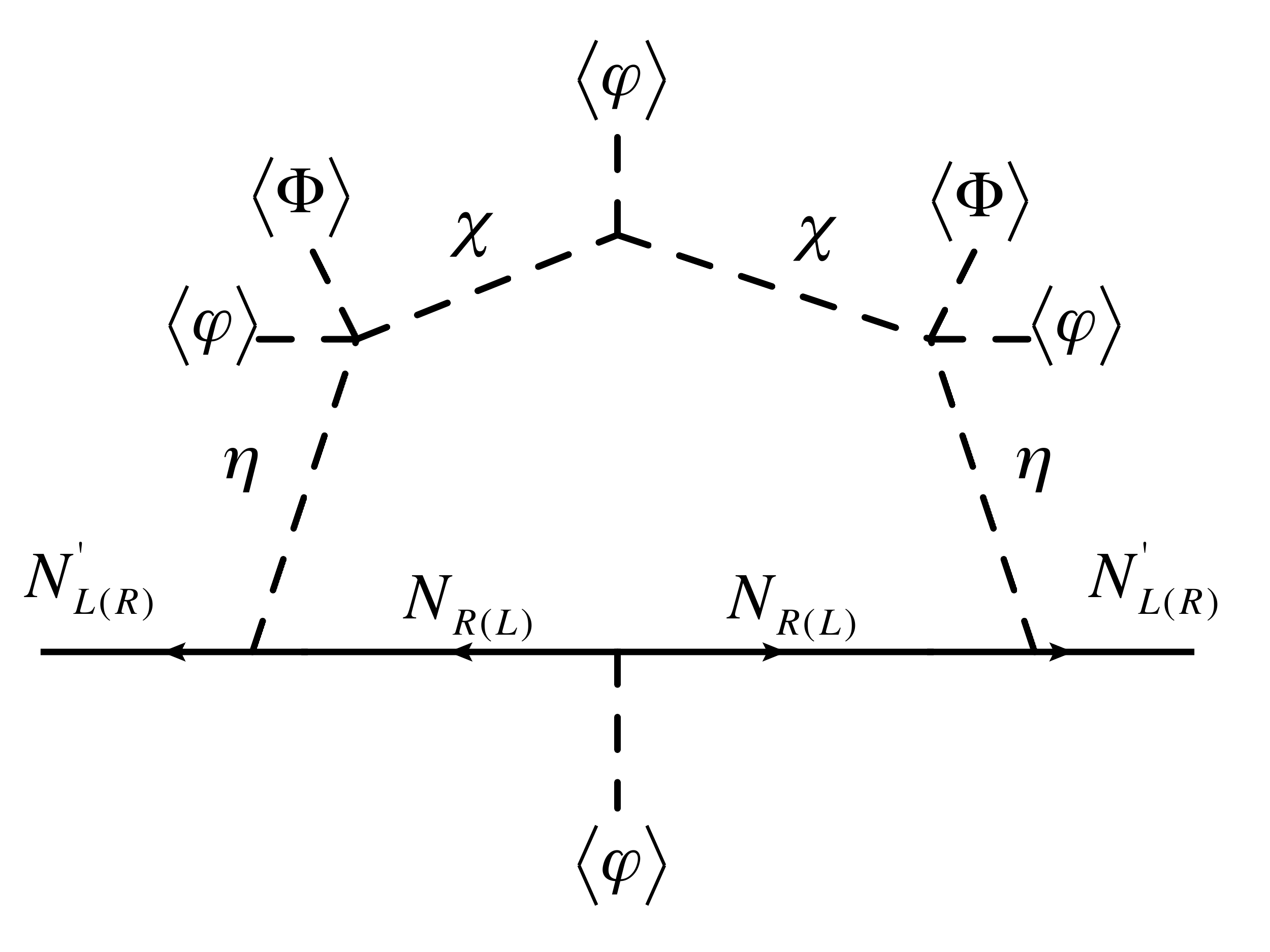}
\caption{The one-loop diagram generating Majorana mass terms of $N'_{L(R)}$.}
\label{fig:diagram}
\end{figure}

We obtain neutral fermion mass terms at tree level, after scalar fields developing VEVs, such that 
\begin{align}
L \supset \bar \nu_L M_D N'_R + \bar N'_L M_{L'} N'_R + \bar{N}_R M_N N_L + \bar N^c_R M_R N_R + \bar N^c_L M_L N_L + h.c.,
\end{align}
where $M_D \equiv y_D v_\varphi/\sqrt{2}$ and $M_{L/R} \equiv y_{\varphi_{L/R}} v_\varphi/\sqrt{2}$.
Then neutrino mass matrices at tree level are obtained as
\begin{equation}
L_{M_\nu}^{\text{tree } } = \frac{1}{2} \left( \begin{array}{c} \bar \nu^c_L \\ \bar N'_R \\ \bar N'^c_L \end{array} \right)^T
\left( \begin{array}{ccc} 0 & M^*_D & 0 \\ M_D^\dagger & 0 & M_{L'}^\dagger \\ 0 & M_{L'}^* & 0 \end{array} \right)
\left( \begin{array}{c} \nu_L \\  N'^c_R \\  N'_L \end{array} \right) + h.c.
\label{Mass_ISS}
\end{equation}
\begin{eqnarray}
L_{M_N}^{\text{tree}} = \frac{1}{2}\left(\begin{array}{c}
\bar{N}^c_L \\
\bar{N}_R
\end{array} \right)^T
\left(
\begin{array}{cc}
2M_L & M^T_N \\
M_N & 2M^\dagger_R \\
\end{array}
\right)
\left(
\begin{array}{c}
N_L \\
N^c_R
\end{array}
\right) + h.c.
\label{Mass_matrix_neutrino}
\end{eqnarray}
The $22$ and $33$ components of the mass matrix in Eq.~\eqref{Mass_ISS} are generated at one-loop level by diagrams in Fig.~\ref{fig:diagram}.
Calculating the diagrams, we obtain one-loop contributions
\begin{eqnarray}
\delta \mu_{22(33)} &=& -\sum_\alpha\frac{{Y_R}_{2(3)\alpha}{Y^T_R}_{\alpha2(3)}}{32 \pi^2} M_{N_\alpha}\int_0^1 dx\Big[ \cos^2 \alpha \log Q^2_{H_1} + \sin^2\alpha \log Q^2_{H_2} \nonumber \\ 
&-& \cos^2\gamma \log Q^2_{A_1} - \sin^2\gamma \log Q^2_{A_2} \Big]
\end{eqnarray}
where $Q_{\phi}^2=x M_{N_\alpha}^2 + (1-x)m_{\phi}^2$ and $M_{N_\alpha}$ is the mass obtained after diagonalizing Eq.~[\ref{Mass_matrix_neutrino}].

Then the inverse seesaw mass matrix becomes
\begin{equation}
L_{M_\nu}^{\text{tree + 1-loop} } = \frac{1}{2} \left( \begin{array}{c} \bar \nu^c_L \\ \bar N'_R \\ \bar N'^c_L \end{array} \right)^T
\left( \begin{array}{ccc} 0 & M^*_D & 0 \\ M_D^\dagger & \delta \mu_{22} & M_{L'}^\dagger \\ 0 & M_{L'}^* & \delta \mu_{33} \end{array} \right)
\left( \begin{array}{c} \nu_L \\  N'^c_R \\  N'_L \end{array} \right) + h.c.
\end{equation}
Since our model has the following hierarchies $\delta \mu_{22(33)}, M_D << M_{L'}$,~\footnote{We will see $M_D/M_{L'}<<1$ is demanded by LFVs in the subsection III~B. $\delta\mu_{22(33)}<< M_{L'}$ would be rather natural, since $\mu_{22(33)}$ is induced at one-loop level while $M_{L'}$ is bare mass. } the inverse seesaw mass matrix can be block diagonalized by unitary matrix $U_{\nu}$ to obtain the mass matrix of the light (active) neutrinos and heavy neutrinos as
\begin{eqnarray}
M_{\nu} &=& -\left(\begin{array}{cc}
M^*_D & 0\\
\end{array}\right)
\left(
\begin{array}{cc}
\delta \mu_{22} & M^\dagger_{L'} \\
M^*_{L'} & \delta \mu_{33} \\
\end{array}
\right)^{-1}
\left(
\begin{array}{c}
M_D^\dagger \\
0\\
\end{array}
\right) \\ \nonumber
&=& M_D^* {M^*_{L'}}^{-1} \delta m_{33} {M^\dagger_{L'}}^{-1} M_{D}^\dagger 
\end{eqnarray}
and 
\begin{eqnarray}
M_{N'}=\left(
\begin{array}{cc}
\delta \mu_{22} & M^\dagger_{L'} \\
M^*_{L'} & \delta \mu_{33}\\
\end{array}
\right)
\end{eqnarray}
such that 
\begin{eqnarray}
U_\nu^T
\left( \begin{array}{ccc} 0 & M^*_D & 0 \\ M_D^\dagger & \delta \mu_{22} & M_{L'}^\dagger \\ 0 & M_{L'}^* & \delta \mu_{33} \end{array} \right) U_\nu
= \left(\begin{array}{cc}
M_{\nu} & 0 \\
0 & M_{N'}\\
\end{array}\right)
\end{eqnarray}
where the unitary matrix $U_\nu$ is given by
\begin{align}
U_\nu = \left[
\begin{array}{cc}
1_{3\times3} & \left(\begin{array}{cc}
M_D^* & 0
\end{array}\right)
\left(\begin{array}{cc}
\delta \mu_{22} & M^\dagger_{L'}\\
M^*_L & \delta \mu_{33} \\
\end{array}\right)^{-1} \\
- \left(\begin{array}{cc}
\delta \mu_{22} & M^\dagger_{L'}\\
M^*_L & \delta \mu_{33} \\
\end{array}\right)^{-1}\left(\begin{array}{c}
M_D^\dagger\\
0\\
\end{array}\right) & 1_{6\times6}
\end{array}
\right] \nonumber \\
= \left[\begin{array}{cc}
1_{3\times3} & \begin{array}{cc}
-M_D^*(M_{L'}^*)^{-1}\delta\mu_{33} (M_{L'}^\dagger)^{-1} & M_D^* (M_{L'}^*)^{-1}
\end{array}\\
\begin{array}{c}
(M_{L'}^*)^{-1}\delta \mu_{33} (M_{L'}^\dagger)^{-1}M_D^\dagger\\
-(M_{L'}^\dagger)^{-1}M_{D}^\dagger
\end{array} & 1_{6\times6}
\end{array}\right]
\end{align}

Using the Cholesky factorization procedure, the symmetric matrix $N={M^*_{L'}}^{-1} \delta m_{33} {M^\dagger_{L'}}^{-1}$ can be written as $N=R R^T$ where $R$ is an upper-right triangle matrix. The light neutrino mass matrix $M_{\nu}$ can be diagonalized by Pontecorvo-Maki-Nakagawa-Sakata (PMNS) matrix $U_{PMNS}$ to obtain the physical masses of the light neutrinos $D_\nu={\rm diag}(m_{\nu_1}, m_{\nu_2}, m_{\nu_3})$
\begin{eqnarray}
D_\nu = U_{PMNS}^T M_{\nu} U_{PMNS} 
\end{eqnarray}
The standard parameterization of $U_{PMNS}$ in terms of mixing angles $\theta_{12},~\theta_{13},~\theta_{23}$, Dirac $CP$ phase $\delta $, and Majorana phases $\alpha_1,~\alpha_2$ is 
\begin{eqnarray}
U_{PMNS}=\left(
\begin{array}{ccc}
c_{12}c_{13}e^{i\alpha_1/2} & s_{12}c_{13}e^{i\alpha_2/2} & s_{13}e^{-i\delta}\\
(-s_{12}c_{23} -c_{12}s_{23}s_{13}e^{i\delta})e^{i\alpha_1/2} & (c_{12}c_{23}-s_{12}s_{23}s_{13}e^{i\delta})e^{i\alpha_2/2} & s_{23}c_{13}\\
(s_{12}s_{23}-c_{12}c_{23}s_{13}e^{i\delta})e^{i\alpha_1/2} & (-c_{12}s_{23}-s_{12}c_{23}s_{13}e^{i\delta})e^{i\alpha_2/2} & c_{23}c_{13}\\
\end{array}
\right)
\end{eqnarray}
where $s(c)_{ij} = \sin(\cos)\theta_{ij}$ and without any loss of generality we can consider $\theta_{ij} \in [0,\pi/2]$.
Following Casas-Ibarra parameterization, we get 
\begin{eqnarray}
M^*_{D} &=& U^*_{PMNS} D_{\nu}^{1/2} \mathcal{O} R^{-1} \nonumber \\
\text{which implies} ~~ y^*_{D} &=& \frac{\sqrt{2}}{v_{\varphi}} U^*_{PMNS} D_{\nu}^{1/2} \mathcal{O} R^{-1}
\end{eqnarray}
where $\mathcal{O}$ is an orthogonal matrix with three independent complex angles.
In our numerical analysis we will consider the perturbative limit $|y_D| \leq \sqrt{4\pi}$. 

The mass matrix for the heavy neutrino $M_{N'}$ is diagonalized perturbatively by an unitary matrix $U_{N'}$ to obtain the masses $D_{N'}={\rm diag}(m_{N'_1},m_{N'_2},m_{N'_3},m_{N'_4},m_{N'_5},m_{N'_6})$ of the six heavy majorana neutrinos
\begin{eqnarray}
D_{N'}=U_{N'}^T M_{N'}U_{N'}=\left(
\begin{array}{cccccc}
m_{E'_i}-\delta {\mu_{22}}^{ii} - \delta \mu_{33}^{ii} & 0  \\ 0 & m_{E'_i}+ \delta {\mu_{22}}^{ii} + \delta \mu_{33}^{ii} \\
\end{array}
\right)
\end{eqnarray} 
In the limit $\delta_{\mu_{22(33)}} \to 0$ we can consider the the heavy neutrinos to be Dirac type fermions. In our case we identify $N'$ flavour state as three heavy Dirac neutrinos.

\subsection{Lepton flavor violation and muon $g-2$}
{
The charged lepton mass term obtained from the Yukawa interaction is 
\begin{eqnarray}
L^{\text{mass}}_{\text{lepton}}=\bar{e}_L M_{l'}e_R + \bar{E'}_L M_{L'}E'_R + \bar{e}_L M_D E'_R + h.c.
\end{eqnarray}
which can be written in matrix form as 
\begin{eqnarray}
L^{\text{mass}}_{\text{lepton}} =
\left(\begin{array}{cc}
\bar{e}_L & \bar{E'}_L\\
\end{array}
\right) 
\left(
\begin{array}{cc}
M_{l'} & M_D \\
0 & M_{L'} \\
\end{array}
\right)
\left(
\begin{array}{c}
e_R\\
E'_R\\
\end{array}
\right) + h.c.
\label{Lepton mass matrix}
\end{eqnarray}
where $M_{l'} = y_l v/\sqrt{2}$.
The mass matrix can be diagonalized by bi-unitary transformation
\begin{eqnarray}
M^{\text{diag}}= U_L^\dagger \left(
\begin{array}{cc}
M_{l'} & M_D \\
0 & M_{L'} \\
\end{array}
\right) U_R
\end{eqnarray}
where $U_L$ and $U_R$ are $6 \times 6$ unitary matrices such that the field transformations are 
\begin{eqnarray}
\left(
\begin{array}{c}
e_L \\
E'_L
\end{array}
\right)
=U_L~l_{L}, \quad
\left(
\begin{array}{c}
e_R \\
E'_R
\end{array}
\right)
=U_R~l_{R}.
\end{eqnarray}
Here we consider the elements of the submatrix $M_D$ to be very small in comparison to diagonal heavy lepton mass matrix $M_{L'}$; $M_D \ll M_{L'}$. 
The smallness of the submatrix $M_D$ is required by the constraints from lepton flavor violating (LFV) processes like $BR(l_i \to l_j \gamma)$. 
In such a case, unitary matrices $U_L$ and $U_R$ are approximated as 
\begin{equation}
\label{eq:UnitaryM}
U_L \simeq \begin{pmatrix} \bm{1} & \bm{0} \\ - \left( M_D M_{L'}^{-1} \right)^\dagger  & \bm{1} \end{pmatrix} \equiv 
\begin{pmatrix} \bm{1} & \bm{0} \\ - \delta U  & \bm{1} \end{pmatrix}, 
\quad U_R \simeq \begin{pmatrix} \bm{1} & \bm{0} \\ \bm{0}  & \bm{1} \end{pmatrix},
\end{equation}
where $\bm{0}$ and $\bm{1}$ are $3\times3$ zero and unity matrices respectively.

To estimate LFV BRs the relevant Yukawa interaction terms are
\begin{eqnarray}
L\supset && \bar{\nu}_i \frac{U^{*}_{PMNS_{ji}} y_D^{j\alpha}}{\sqrt{2}}N'_\alpha(\sin\beta h + \cos\beta H) + \bar{N'_\alpha} \frac{y_D^{*j\alpha} U_{PMNS_{ji}}}{\sqrt{2}}P_L \nu_i (\sin\beta h + \cos\beta H) \nonumber \\
&+&
 \bar{e}_i \frac{y_D^{i\alpha}}{\sqrt{2}}E'_\alpha(\sin\beta h + \cos\beta H) + \bar{E'_\alpha} \frac{y_D^{*i\alpha}}{\sqrt{2}}P_L e_i (\sin\beta h + \cos\beta H), \nonumber \\
 &+&  \frac{i}{\sqrt{2}} y_D^{i \alpha}  \bar e_{i} P_R E'_{\alpha} G_{Z'} + h.c., 
\end{eqnarray}  
where summation over repeated index is implied. 
In this Lagrangian the light neutrinos $\nu_i$ are three Majorana neutrinos and $N'_\alpha$ are three heavy Dirac neutrinos.  
Note that we write interaction among $e^i$, $E^\alpha$ and NG boson $G_{Z'}$ to calculate muon $g-2$ in Feynman gauge.
Flavor violating $Z'$ interaction terms are also given by
\begin{align}
\label{eq:LFVZp}
L\supset -x g'' \left[ \delta U_{\alpha i} \bar E'_\alpha \gamma_\mu P_L e_i + (\delta U^\dagger)_{i \alpha} \bar e_i \gamma_\mu P_L E'_\alpha \right] Z'^\mu
+ x g'' (\delta U^\dagger \delta U)_{ij} \bar e_i \gamma_\mu P_L e_j Z'^\mu,
\end{align}
where $\delta U$ is $3\times3$ matrix given in Eq.~\eqref{eq:UnitaryM}.
From the interaction we obtain $Z' \to \ell_i \bar \ell_j$ decay width  for vanishing kinetic mixing such that 
\begin{align}
\Gamma_{Z' \to \ell_i \bar \ell_j} = & \frac{|x g'' (\delta U^\dagger \delta U)_{ij}|^2}{24 \pi} m_{Z'} \lambda^{\frac12}\left(1,\frac{m_{\ell_1}^2}{m_{Z'}^2},\frac{m_{\ell_2}^2}{m_{Z'}^2} \right)
  \\ 
 & \times \left[ 
 1 - \frac{m_{\ell_1}^2}{m_{Z'}^2} - \frac{m_{\ell_2}^2}{m_{Z'}^2} 
 + \sqrt{  \left(\lambda \left(1,\frac{m_{\ell_1}^2}{m_{Z'}^2},\frac{m_{\ell_2}^2}{m_{Z'}^2} \right) + \frac{4  m_{\ell_1}^2}{m_{Z'}^2} \right)  \left(  \lambda \left(1,\frac{m_{\ell_1}^2}{m_{Z'}^2},\frac{m_{\ell_2}^2}{m_{Z'}^2} \right)+ \frac{4m_{\ell_2}^2}{m_{Z'}^2} \right) }  \right], \nonumber
\end{align}
where $\lambda(1,x,y) \equiv 1 + x^2+ y^2 -2x -2y -2xy$.
If light $Z'$ can only decay through this lepton mixing effect its lifetime will be long; for example $\tau_{Z'} \sim 5 \times 10^{-4} s$ for $|(\delta U^\dagger \delta U)_{22}| = 10^{-8}$, $g'' \sim 0.1$ and $m_{Z'} = 1$ GeV.

Considering the one-loop diagram with scalars $h,~H$ and leptons $E'_\alpha$ running in the loop, we obtain~\cite{Lindner:2016bgg, Baek:2016kud} 
\begin{eqnarray}
BR(l_i \to l_j \gamma)= \frac{3(4\pi)^3\alpha_{em}}{4G_F^2}\Big[ |A^M_{ji}|^2 + |A^E_{ji}|^2 \Big]
BR(\l_i \to l_j\nu_i\bar{\nu}_j) \label{eq:LFV}
\end{eqnarray}
where 
\begin{eqnarray}
A^M_{ji} &=& \frac{m_i + m_j}{16\pi^2 m_i}\sin^2\beta \sum_\alpha F(m_{E'_{\alpha}},m_h)y^{j\alpha}_D y^{*i\alpha}_D   \nonumber \\ 
&+& \frac{m_i + m_j}{16\pi^2 m_i}\cos^2\beta \sum_\alpha F(m_{E'_{\alpha}},m_H)y^{j\alpha}_Dy^{*i\alpha}_D \nonumber \\
&-& \frac{m_i + m_j}{8 \pi^2 m_i} x^2 g''^2 \sum_\alpha \hat{F}(m_{E'_{\alpha}},m_{Z'})  \delta U_{\alpha j} (\delta U^\dagger)_{i \alpha}  \nonumber \\
&+& \frac{m_i + m_j}{16\pi^2 m_i} \sum_\alpha F(m_{E'_{\alpha}},m_{Z'})y^{j\alpha}_Dy^{*i\alpha}_D \nonumber \\
A^E_{ji} &=& \frac{m_i - m_j}{16\pi^2 m_i}\sin^2\beta \sum_\alpha F(m_{E'_{\alpha}},m_h)y^{j\alpha}_D y^{*i\alpha}_D   \nonumber \\ 
&+& \frac{m_i - m_j}{16\pi^2 m_i}\cos^2\beta \sum_\alpha F(m_{E'_{\alpha}},m_H) y^{j\alpha}_D y^{*i\alpha}_D  \nonumber \\
&-& \frac{m_i - m_j}{8 \pi^2 m_i} x^2 g''^2 \sum_\alpha \hat{F}(m_{E'_{\alpha}},m_{Z'})  \delta U_{\alpha j} (\delta U^\dagger)_{i \alpha} \nonumber \\ 
&+& \frac{m_i - m_j}{16\pi^2 m_i} \sum_\alpha F(m_{E'_{\alpha}},m_{Z'}) y^{j\alpha}_D y^{*i\alpha}_D,  
\end{eqnarray}
where the last two terms in the RHS are respectively contributions from $Z'$ (transverse mode) and NG boson, and loop integration factor is 
\begin{eqnarray}
F(m_a,m_b) &=& \frac{m_a^6 + 2 m_b^6 - 6 m_a^4 m_b^2 + 3 m_a^2 m_b^4 + 12 m_a^2 m_b^4 \log(\frac{m_a}{m_b})}{12(m_a^2 - m_b^2)^4}, \nonumber \\
\hat{F}(m_a,m_b) &=& \frac{5 m_a^6 + 4 m_b^6 - 9 m_a^2 m_b^4 -12 m_a^2 m_b^2 (2 m_a^2 - m_b^2) \log(\frac{m_a}{m_b})}{12(m_a^2 - m_b^2)^4}, 
\end{eqnarray}
%
 Note that the second term of RHS in Eq.~\eqref{eq:LFVZp} also contributes to $\ell_i \to \ell_j \gamma$ process through $Z'$ loop.
However this effect is suppressed since we take $\delta U = -(M_D/M_{L'})^\dagger$ to be very small.
In our numerical calculation, we use explicit form of Eq.~\eqref{eq:LFV} to impose constraints from the LFV decay BRs. 
In addition we consider $\mu$-$e$ conversion through LFV interaction associated with $Z$ boson induced by the lepton mixing effect~\cite{Crivellin:2020ebi}. 
The effective interaction for $\mu$-$e$ conversion is written as 
\begin{align}
& L_{eff} = \sum_{q=u,d} (C^{LL}_q O_q^{LL} + C^{LR}_q O_q^{LR}) + (L \leftrightarrow R) +h.c. \ , \\
& O_q^{XX} = (\bar e \gamma^\mu P_X \mu)(\bar q \gamma_\mu P_X q), \quad C_q^{XX} = \Gamma^{ X}_{e \mu} \frac{1}{m_Z^2} \Gamma_{qq}^{X},
\end{align}
where $X = L, R$ and $\Gamma^X_{qq(e\mu)}$ is $Z$ couplings given by
\begin{align}
& \Gamma_{uu}^{L} = - \frac{g}{c_W} \left( \frac{1}{2} -\frac{2}{3} s_W^2 \right), \quad \Gamma_{uu}^R = \frac{2}{3} \frac{g s_W^2}{c_W}, \\
& \Gamma_{dd}^{L} = - \frac{g}{c_W} \left( -\frac{1}{2} +\frac{1}{3} s_W^2 \right), \quad \Gamma_{dd}^R = -\frac{1}{3} \frac{g s_W^2}{c_W}, \\
& \Gamma_{e\mu}^{L} \simeq - \frac{g}{c_W} \left( -\frac{1}{2} + s_W^2 \right) \frac{ M_{D_{ea}} M^\dag_{D_{a\mu}}}{M^2_{L'_a}} , \quad \Gamma_{e\mu}^R \simeq 0, 
\end{align}
with $s_W(c_W) = \sin \theta_W (\cos \theta_W)$.
Then $\mu$--$e$ conversion ratio is given by~\cite{Crivellin:2017rmk,Cirigliano:2009bz,Kitano:2002mt}
\begin{equation}
\Gamma^N_{\mu \to e} = 4 m_\mu^5 \left| \sum_{q=u,d} \left( C_q^{RL} + C_q^{RR} \right) \left( f_p^q V_N^p + f_n^q V_N^n \right)  \right|^2 + (L \leftrightarrow R),
\end{equation}
where $f_p^u =2$, $f_n^u =1$, $f_p^d =1$, $f_n^d =2$ and $V_N^{p(n)}$ is the overlap integrals for nuclear $N$. 
For gold, we obtain $V_{Au}^{p(n)} = 0.0974(0.146)$~\cite{Kitano:2002mt}. 
We need to normalize the conversion ratio by capture rate $\Gamma_{Au}^{\rm capt} = 8.7 \times 10^{-18}$ GeV~\cite{Suzuki:1987jf}, and the constraint is given by~\cite{Bertl:2006up}
\begin{equation}
\frac{\Gamma^{Au}_{\mu \to e}}{\Gamma_{Au}^{\rm capt}} < 7.0 \times 10^{-13}.
\end{equation}
We then obtain the constraint on lepton mixing such that
\begin{equation} 
\frac{ M_{D_{ea}} M^\dag_{D_{a\mu}}}{M^2_{L'_a}} \lesssim 2.8 \times 10^{-7}
\end{equation}
The constraint is weaker than $\mu \to e \gamma$ but it will be tested in future experiments with more precision.
Hence the lepton mixing should be very small, and for a good approximation we can consider no lepton mixing when we consider SM lepton interactions.

}

\noindent
{\it Muon anomalous magnetic moment} \\
It is known as hint of physics beyond the SM since observed value is deviated from the SM prediction. 
The deviation from the SM prediction $\Delta a_\mu$ is  about 3.7$\sigma$~\cite{Bennett:2006fi, Aoyama:2020ynm};
\begin{align}
\Delta a_\mu=(2.706\pm0.726)\times 10^{-9}.~\label{eq:mg2}
\end{align}
It suggests that new physics would be needed.
 Our new contribution to the muon $g-2$ is obtained to be
\begin{eqnarray}
\Delta a_\mu &=& \frac{m_{\mu}^2}{8\pi^2}\sin^2\beta \sum_\alpha F(m_{E'_\alpha},m_h) y_D^{2\alpha} y_D^{*2\alpha} \nonumber \\
&+& \frac{m_{\mu}^2}{8\pi^2}\cos^2\beta \sum_\alpha F(m_{E'_\alpha},m_H) y_D^{2\alpha} y_D^{*2\alpha} \nonumber \\
&-& \frac{m_{\mu}^2}{4 \pi^2}  x^2 g''^2 \sum_\alpha \hat{F} (m_{E'_\alpha},m_{Z'}) \delta U_{\alpha 2} (\delta U^\dagger)_{2 \alpha} \nonumber \\
&+& \frac{m_{\mu}^2}{8\pi^2}  \sum_\alpha F(m_{E'_\alpha},m_{Z'}) y_D^{2\alpha} y_D^{*2\alpha},
\end{eqnarray}
where the last two terms in the RHS are respectively contributions from $Z'$ (transverse mode) and NG boson. 

\noindent
{\it Flavor violating $\ell_i \to \ell_j Z'$ decay} \\
We also have this kind of LFV process induced by the second term of Eq.~\eqref{eq:LFVZp}.
For $m_{Z'} < m_i - m_j$, decay width is given by
\begin{align}
& \Gamma(\ell_i \to \ell_j Z') = x^2 g''^2 (\delta U^\dagger \delta U)^2_{ij} \frac{k}{8 \pi m_i}  \left( \sqrt{k^2 + m_j^2} + \frac{1}{m_{Z'}^2} (m_i^2 - m_j^2 -m^2_{Z'})\sqrt{k^2 + m^2_{Z'}} \right), \\
& k = \frac{m_i}{2} \lambda^{\frac12}\left(1, \frac{m_j^2}{m^2_i}, \frac{m^2_{Z'}}{m^2_i}  \right). \nonumber 
\end{align}
Then $Z'$ decays into leptons through the mixing effect giving three charged lepton final states.
Thus when these modes are kinematically allowed there are strong constraint for $\delta U = M_D^\dagger/M_{L'}$.
For $m_{Z'} < m_\mu$, it is constrained by $BR(\mu \to 3 e) \lesssim 10^{-12}$~\cite{Bellgardt1988} which results in 
\begin{equation} 
\left| g'' \frac{ M_{D_{ea}} M^\dag_{D_{a\mu}}}{M^2_{L'_a}} \right| \lesssim 9 \times 10^{-15}
\end{equation}
where we used $m_{Z'} = 0.05$ GeV as a reference value.
For $m_\mu < m_{Z'} < m_\tau$, it is constrained by , for example $BR(\tau \to 3 \mu) \lesssim 2 \times 10^{- 8}$~\cite{Hayasaka2010}, which results in 
\begin{equation} 
\left| g'' \frac{ M_{D_{\mu a}} M^\dag_{D_{a\tau}}}{M^2_{L'_a}} \right| \lesssim 4 \times 10^{-10}
\end{equation}
where we used $m_{Z'} = 0.5$ GeV as a reference value.
Thus we find that strong constraints are imposed from $\ell_i \to \ell_j Z'$ when these decay modes are kinematically allowed.
In our analysis, we consider $Z' > m_{\tau}$ to avoid these strong LFV constraints.

\subsection{Leptonic $Z$ boson decays}
The $Z_\mu $ to the leptonic (charged and uncharged) interaction terms in the limit $\epsilon \to 0$ i.e. $(\sin\theta \to 0)$ is given by~\cite{Chiang:2017tai, Kumar:2020web, Nomura:2019btk}
\begin{eqnarray}
L &\supset& \frac{g}{c_W}\Big[ \bar{l}\gamma_\mu \Big(\frac{P_L}{2} - s_W^2\Big)l + \bar{E'} \gamma_\mu \Big(\frac{1}{2} -s_W^2\Big)E' \Big]Z^\mu
\nonumber \\
&-& \frac{g}{c_W} \Big[ \bar{\nu} \gamma_\mu \frac{P_L}{2}\nu + \bar{N'}\gamma_\mu \frac{1}{2}N' \Big]Z^\mu 
\end{eqnarray}
where the sine of the Weinberg angle is $s_W \sim 0.23$. The equation is written in the mass basis where $\nu$ are the three light Majorana neutrinos and $N'$ are three heavy Dirac neutrinos.
The deviation of $Z$ boson branching ratios from SM prediction up to 1-loop level is given by 
\begin{align}
\Delta {\rm BR} (Z \to f_{i}\bar{f_{j}}) 
&= \frac{\Gamma (Z \to f_{i}\bar{f_{j}})_{\rm SM+ New} - \Gamma (Z \to f_{i}\bar{f_{j}})_{\rm SM}}{\Gamma^{tot}_Z} \nonumber \\
&\approx 
\left\{~\begin{array}{cc}
 \frac{\Gamma(Z\to f_i\bar f_j)_{\rm SM+New} -  \Gamma(Z\to f_i\bar f_j)_{\rm SM}}{\Gamma^{\text{tot}}_Z}  & (i =j)\\
\frac{ \Gamma(Z\to f_i\bar f_j)_{\rm  New}} {\Gamma^{\text{tot}}_Z}  & (i \neq j)\\
\end{array}\right.,
\end{align} 
where $\Gamma_Z^{\rm{tot}}=2.4952 \pm 0.0023$~\cite{Tanabashi:2018oca}.
Then our new contributions for charged-lepton and neutrino final modes are respectively given by  
\begin{eqnarray}
\Delta {\rm BR}(Z \to l_i \bar{l_j}) &=& \frac{m_Z g^2}{12 \pi \Gamma_Z^{\rm{tot}} c^2_W}
\Big[ \frac{|B^l_{ij}|^2}{2} - s^2_W  \delta_{ij} Re [B^{l*}_{ij}] 
- \Big( - \frac{s_W^2}{2}+ \frac{1}{8}\Big)\delta_{ij}\Big] ,\nn  \\
\Delta {\rm BR}(Z \to \nu_i \bar{\nu_j})&=& \frac{m_Z g^2}{24 \pi \Gamma_Z^{\rm{tot}} c^2_W}
\Big[ |B^\nu_{ij}|^2 -\frac{\delta_{ij}}{4} \Big],
\end{eqnarray}
with  
\begin{eqnarray}
&&B^l_{ij} = \frac{\delta_{ij}}{2} - \frac{1}{2}\sum_{\alpha} \frac{y_D^{i\alpha}y_D^{*j\alpha}}{(4\pi)^2} G^l(m_{E'_\alpha},m_{h})\sin^2\beta - \frac{1}{2}\sum_\alpha \frac{y_D^{i\alpha}y_D^{*j\alpha}}{(4\pi)^2} G^l(m_{E'_\alpha},m_{H})\cos^2\beta  , \nonumber \\
&& B^\nu_{ij} = \frac{\delta_{ij}}{2} + \frac{1}{2}\sum_{\alpha,k,l} \frac{U^*_{PMNS_{ki}} U_{PMNS_{lj}} y_D^{k\alpha}y_D^{*l\alpha}}{(4\pi)^2} G^{\nu}(m_{N'_\alpha},m_{h})\sin^2\beta \nonumber \\
&& \hspace{1.75cm} + \frac{1}{2}\sum_{\alpha,k,l} \frac{U^*_{PMNS_{ki}} U_{PMNS_{lj}} y_D^{k\alpha}y_D^{*l\alpha}}{(4\pi)^2} G^{\nu}(m_{N'_\alpha},m_{H})\cos^2\beta , \nonumber \\
&&G^l(m_a,m_b) = -\Big( -\frac{1}{2} + s^2_W \Big)^2 H_1(m_a,m_b) + \Big( -\frac{1}{2} + s^2_W \Big) H_2(m_a,m_b) \nonumber \\
&&G^{\nu}(m_a,m_b) = -\frac{1}{2}H_1(m_a,m_b) - \frac{1}{2}H_2(m_a,m_b) ,
\nonumber \\&& H_1(m_a,m_b) = \frac{m_b^4 - 4m_a^2 m_b^2 + 3 m_a^4 -4m_b^2(m_b^2 - 2 m_a^2)\log(m_b) - 4 m_a^4 \log(m_a)}{4(m_a^2 -m_b^2)^2} ,\nn \\
&& H_2(m_a,m_b) = m_a^2 \Big[ \frac{m_a^2 - m_b^2 + 2 m_b^2 \log(\frac{m_b}{m_a})}{(m_a^2 + m_b^2)^2} \Big].
\end{eqnarray}

\subsection{Numerical Analysis}

In our numerical analysis we consider the Yukawa couplings $|y_D^{1\alpha}|\in [10^{-8},10^{-3}]$, $|y_D^{2\alpha}|\in [10^{-3},1.0]$ and $|y_D^{3\alpha}|\in [10^{-6},10^{-1}]$, $m_H \in [1,10]$ GeV, $m_{E'_\alpha} \in [100,1500]$ GeV, $m_{N'_\alpha} \in [100,1500]$ GeV and $\sin\beta \sim 10^{-4}$. The allowed parameter spaces are obtained which satisfy the neutrino data of recent global fit by NuFIT 5.0~\cite{Esteban:2020cvm}. For simplicity we consider normal ordering and the Majorana phases as free parameters in the range $[0,2 \pi]$. 
\begin{align}
& s^2_{12} \in [0.269,0.343], ~~~ s^2_{23} \in [0.415,0.616], ~~~ s^2_{13}\in [0.02032,0.02410],~~ \delta \in \left[ \frac{2}{3}\pi, 2\pi \right], \nonumber \\
& \Delta m_{21}^2 \in [6.82,8.04]\times 10^{-5} {\rm eV} ~~ \text{and} ~~ \Delta m_{31}^2 \in [2.435,2.598]\times 10^{-3} {\rm eV}.   
\end{align}   
 \begin{figure}[tb]
\includegraphics[width=10cm]{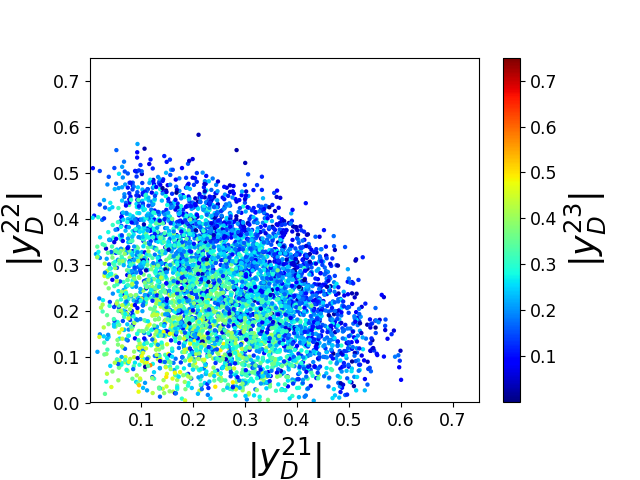}
\caption{Allowed region of $|y_D^{21}|$, $|y_D^{22}|$ and $|y_D^{23}|$}
\label{fig:allowedyD}
\end{figure}
Under these parameter space, we have randomly selected and show plots to satisfy all the constraints as discussed above.
Fig.~\ref{fig:allowedyD} shows the allowed spaces for relevant components of the Dirac Yukawa couplings $|y^{{21}}_{D}|$, $|y^{{22}}_{D}|$ and $|y^{{23}}_{D}|$,
which are characterized by Casas-Ibarra parametrization and restricted by leptonic $Z$ boson decays as well as LFVs. It implies that these components are 0.6 at most that would be reasonable in our model.

\begin{figure}[tb]
\includegraphics[width=80mm]{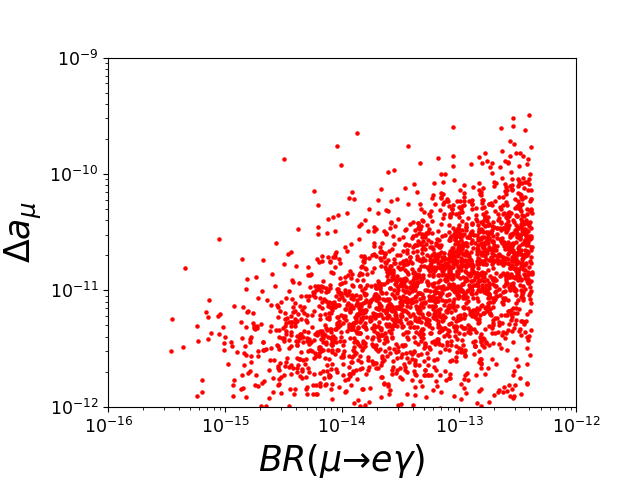}
\includegraphics[width=80mm]{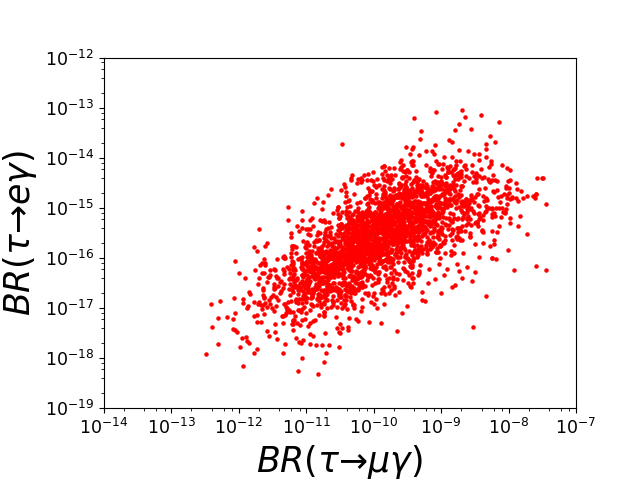}
\caption{Allowed region for LFVs and the correlation to muon $g-2$. The left one shows muon $g-2$ in terms of BR($\mu\to e\gamma)$, while  the right one BR($\tau\to e\gamma)$ in terms of BR($\tau\to \mu\gamma)$.}
\label{fig:muon g-2 and LFV}
\end{figure}
Fig.~\ref{fig:muon g-2 and LFV} shows the possibly allowed regions for muon $g-2$, BR($\mu\to e\gamma)$ in the left figure, and  BR($\tau\to e\gamma)$ and  BR($\tau\to \mu\gamma)$ in the right one.
In the left one, new contribution to the muon $g-2$ reaches at $3\times 10^{-10}$ at around the the upper bound on BR($\mu\to e\gamma)=4.2\times 10^{-13}$. Even though the scale of muon $g-2$ is smaller than the expected value in Eq.~\eqref{eq:mg2} by 1 magnitude, it will still be verifiable to be tested in near future.
In the right one,   BR($\tau\to e\gamma)$ is much smaller than the current upper bound while BR($\tau\to \mu\gamma)$ reaches the current upper bound. Therefore, BR($\tau\to \mu\gamma)$ could also be tested in near future.   

\section{Dark matter physics}

\begin{figure}[t]
\includegraphics[width=80mm]{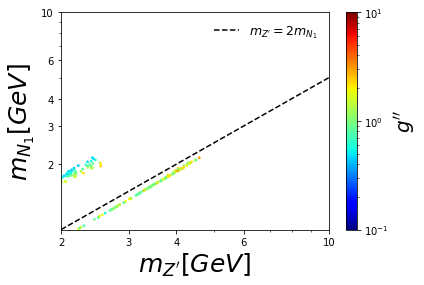}
\includegraphics[width=80mm]{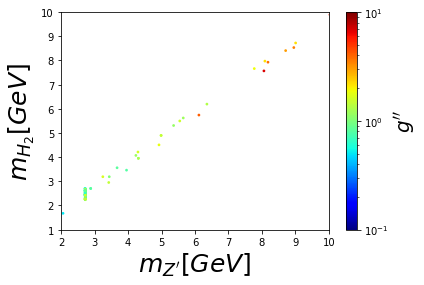}
\caption{Allowed regions of parameter space satisfying the observed DM relic density constraint. Left: The DM candidate is $N_1$ and Right: The DM candidate is $H_2$.}
\label{fig:allowed_relic_density}
\end{figure}

For the dark matter (DM) analysis we consider two scenarios, fermionic DM scenario and the scalar DM scenario.
{\it In the fermionic DM scenario}, we consider the lightest of the $Z_2$ odd Majorana fermions $N_{1,..,6}$ obtained by diagonalizing Eq.\eqref{Mass_matrix_neutrino} as the DM candidate. For simplicity we consider the mass ordering $m_{N_1}<m_{N_2}<m_{N_3}<m_{N_4}<m_{N_5}<m_{N_6}$ and therefore the lightest DM candidate is $N_1$. 
For numerical analysis we consider  $m_{N_1} \in [1,10]$ GeV, $m_{Z'} \in [2,10]$ GeV, the gauge kinetic mixing parameter $\epsilon \sim 10^{-4}$ and the $U(1)_X$ charge, $x=1$, $m_H \in [1,10]$ GeV, $v_{\varphi} \in[1,5]$ GeV and $\sin\beta \sim 10^{-4}$. The relic density is computed using the public code Micromegas-5.2.4~\cite{Belanger:2014vza} implementing our model. The left panel of Fig.~\ref{fig:allowed_relic_density} shows the allowed region of $m_{Z'}-m_{N_1}$ space with $g''$ in colour bar satisfying the observed relic density;  $0.11 < \Omega h^2 < 0.13$. Above the dashed line, the dominant contribution to $\langle \sigma v \rangle$ comes from the $N_1 N_1 \to Z' Z'$ channel. Below the dashed line, the dominant contribution to $\langle \sigma v \rangle$ comes from $N_1 N_1 \to Z'\to f_{SM} \bar{f}_{SM}$ processes. Along the dashed line slightly below it, the cross sections are enhanced due to resonance which leads to a sharp fall in the relic density \cite{Nayak:2017dwg}.

{\it For the scalar DM scenario},
we have two candidates; $H_1$ or $H_2$, where one can select the lighter one between CP even and CP odd bosons. When the mixing is not so large, $H_1$ is $\eta_R$ dominant and $H_2$ is $\chi_R$ one.
In case of $H_1$, the detailed behavior of DM has been known unless the Yukawa interactions $y_{L,R}$ are dominant. Below 100 GeV, a solution has at the pole of half of DM Higgs mass $\sim63$ GeV. Above 100 GeV,
another solution is at around 534 GeV that is induced by gauge interactions in kinetic terms.
 In details, see, e.g., ref.~\cite{Hambye:2009pw}. However since our focus region is that the DM mass is less than 1 GeV, these solution is not valid.
Even though we have new channels via $H_1-A_1-Z'$ interactions, the DM mass wouldn't be below 1 GeV because of the electroweak precision test. 
Thus, we consider $H_2$ as the lightest $Z_2$ odd particle. 
For numerical analysis we consider  $m_{H_2} \in [1,10]$ GeV, $m_{Z'} \in [2,10]$ GeV and the remaining parameters same as above. The right panel of Fig.~\ref{fig:allowed_relic_density} shows the allowed region of $m_{Z'}-m_{H_2}$ space with $g''$ in color bar satisfying the observed relic density. The dominant contribution to the relic density comes from $H_2 H_2 \to Z' \to f_{SM} \bar{f}_{SM}$ channel.

\section{Summary and discussion}
We have proposed an inverse seesaw scenario under hidden $U(1)$ gauge symmetry, having natural hierarchies among mass scales of neutral fermions.
The hierarchies are derived from theory and experimental constraints. The theoretical aspect tells us $\delta\mu_{22,33}<<M_{L'}$ since  $\delta\mu_{22,33}$ is induces at one-loop level, while $M_{L'}$ is bare mass.
The experimental side requests that $M_D<<M_{L'}$ since $M_D$ is highly suppressed by the LFVs such as $\mu\to e\gamma$. In order to induce small $\delta \mu_{22,33}$, we introduce exotic fermions and bosons that provide us additional intriguing phenomenologies such as muon anomalous magnetic moment, dark matter candidate as well as LFVs and leptonic $Z$ decays. We have analyzed these phenomenologies including neutrino oscillation data numerically, and shown allowed region. Here, Dirac Yukawa coupling $y_D$ possesses the information of neutrino data derived by modified Casas-Ibarra parametrization.
We have found that our new contribution to the muon $g-2$ is about $3\times 10^{-10}$ which is smaller than 1 order magnitude, but it would still be verifiable in near future experiments. 
BR$(\mu\to e\gamma)$ and BR$(\tau\to \mu\gamma)$ would be tested soon since these upper bounds reaches at the current experimental bounds. Finally, we have discussed the DM candidate in both the cases of fermion and boson,
where we have focussed on rather lighter range that is equal or less than 1 GeV. The dominant contributions originate from interactions of hidden gauge sector, and we have found allowed ranges for both the case as can be seen in Fig.~\ref{fig:allowed_relic_density}.

In our scenario, we consider light $Z'$ and extra scalar $H$ since it is preferred to provide sizable muon $g-2$ relaxing tension between experimental data and theoretical prediction.
In such a case the SM Higgs boson can decay into $HH$ and $Z'Z'$ mode where $H$ further decays into $Z'Z'$ and $Z'$ decays into charged leptons via lepton mixing effect.
Thus we would have multi-lepton signals from rare Higgs decay as $h \to Z' Z' \to 4 \ell$ and/or $h \to HH \to 4Z' \to 8 \ell$ which can be distinguishable signature of our scenario.
Interestingly, leptonic $Z'$ decays includes LFV modes.
We need detailed simulation analysis to estimate discovery potential of these signals since $H$ and $Z'$ are light and some final state particles are collimated.
Therefore we leave the analysis in future work.

Before closing our paper, we will briefly mention collider phenomenologies.
In our scenario, extra charged particles $\eta^\pm$ and $E^\pm$ can be produced by electroweak interactions at the LHC.
Inert charged scalar $\eta^\pm$ dominantly decays into $W^{\pm (*)} H_{1,2}$ assuming $E^\pm$ has heavier mass; $H_{1,2}$ would be DM or it decays into the states including DM with other particles.
If $\eta^\pm$ decays into $W^{\pm(*)} DM$ the situation is the same as inert Higgs doublet scenario, and it is well studied, e.g. in ref.~\cite{Belyaev:2018ext}.
Exotic lepton $E^\pm$ can decay into $\ell^\pm H$ and $\eta^\pm N_i$ where $H$ decays into $Z'Z'$ and $\eta^\pm$ decays as described above.
Thus new signals from these exotic charged particles provide us complicated final states with many particles and analysis of them is beyond the scope of this paper.
Analysis of collider signals would be given in future work. 

\vspace{0.5cm}
\hspace{0.2cm} 
\begin{acknowledgments}
The work of H.O. and P.S. was supported by the Junior Research Group (JRG) Program at the Asia-Pacific Center for Theoretical
Physics (APCTP) through the Science and Technology Promotion Fund and Lottery Fund of the Korean Government and was supported by the Korean Local Governments-Gyeongsangbuk-do Province and Pohang City. The authors would like to thank Dr.~Arindam Das for his fruitful discussions.
H.O. is sincerely grateful for all the KIAS members.
\end{acknowledgments}

\appendix

\section{Couplings in scalar potential}

The parameters in the scalar potential can be written by scalar masses and mixings:
\begin{eqnarray}
\lambda_\Phi &=& \frac{m^2_h\cos^2\beta + m_H^2\sin^2\beta}{2v^2} \nonumber \\
\lambda_\varphi &=& \frac{m_h^2\sin^2\beta + m_H^2\cos^2\beta}{2v_\varphi^2} \nonumber \\
\lambda_{\Phi\varphi} &=& -\frac{(m_H^2-m_h^2)\cos\beta\sin\beta}{v v_\varphi} \nonumber \\
\lambda_0 &=& -\frac{2(m^2_{H_2} - m^2_{H_1})\cos\alpha\sin\alpha}{v v_\varphi} \nonumber \\
&=& \frac{2(m^2_{A_2} - m^2_{A_1})\cos\gamma\sin\gamma}{v v_\varphi} \nonumber \\
\tilde{\lambda}_{\Phi\eta} &=& -\frac{2(m_{\eta^\pm}^2 -m_{H_1}^2\cos^2\alpha - m_{H_2}^2\sin^2\alpha)}{v^2} \nonumber \\
&=& -\frac{2(m_{\eta^\pm}^2 -m_{A_1}^2\cos^2\gamma - m_{A_2}^2\sin^2\gamma)}{v^2} \nonumber\\
m_{\chi^\prime}^2 &=& m_{A_1}^2 \sin^2\gamma + m_{A_2}^2 \cos^2\gamma \nonumber \\
m_{\chi^{\prime\prime}}^2 &=& m_{H_1}^2 \sin^2\alpha + m_{H_2}^2 \cos^2\alpha \nonumber \\
\mu &=& \frac{m_{\chi^{\prime\prime}}^2 - m_{\chi^\prime}^2}{2\sqrt{2}v_\varphi} \nonumber \\
m_{\chi}^2 &=& \frac{1}{2}\Big[ m_{\chi^\prime}^2 + m_{\chi^{\prime\prime}}^2 -v^2\lambda_{H\chi} -v_{\varphi}^2\lambda_{\varphi\chi} \Big].
\end{eqnarray}
Thus the independent parameters are $m_h, m_H, m_{H_1}, m_{H_2}, m_{A_1}, m_{A_2}, m_{\eta^\pm}, v, v_{\varphi}, \cos\beta,$\\ $\lambda_{\Phi \chi},\lambda_{\varphi\chi},\lambda_{\Phi \eta}, \lambda_{\eta\varphi}, \lambda_{\eta\chi}, \lambda_{\eta}, \lambda_{\chi}$.

\end{document}